%% file: main.tex
\documentclass[conference]{IEEEtran}

\input{includes}

\IEEEoverridecommandlockouts
\usepackage{cite}
\usepackage{amsmath,amssymb,amsfonts}
\usepackage{graphicx}
\usepackage{textcomp}
\usepackage{xcolor}
\def\BibTeX{{\rm B\kern-.05em{\sc i\kern-.025em b}\kern-.08em
    T\kern-.1667em\lower.7ex\hbox{E}\kern-.125emX}}

\newcommand{\tv}[1]{\footnote{{\bf Thiemo: #1}}}

\begin{document}

\title{\ours: A Unified Benchmark Suite for Byzantine Attacks and Defenses in Federated Learning
}

    
\author{
\IEEEauthorblockN{
Shenghui LI\IEEEauthorrefmark{1}, 
Edith C.-H. Ngai\IEEEauthorrefmark{2},
Fanghua Ye\IEEEauthorrefmark{3},
Li Ju\IEEEauthorrefmark{1},
Tianru Zhang\IEEEauthorrefmark{1},
Thiemo Voigt\IEEEauthorrefmark{1}\IEEEauthorrefmark{4}}
\IEEEauthorblockA{\IEEEauthorrefmark{1}Uppsala University, Uppsala, Sweden. \emph{\{shenghui.li,li.ju,tianru.zhang\}@it.uu.se}}
\IEEEauthorblockA{\IEEEauthorrefmark{2}The University of Hong Kong, Hong Kong, China. \emph{chngai@eee.hku.hk}}
\IEEEauthorblockA{\IEEEauthorrefmark{3}University College London, London, UK. \emph{fanghua.ye.19@ucl.ac.uk}}
\IEEEauthorblockA{\IEEEauthorrefmark{4}Research Institutes of Sweden, Stockholm, Sweden. \emph{thiemo.voigt@angstrom.uu.se}}
}

\maketitle

\begin{abstract}
Federated learning (FL) facilitates distributed training across different IoT and edge devices, safeguarding the privacy of their data. The inherent distributed structure of FL introduces vulnerabilities, especially from adversarial devices aiming to skew local updates to their advantage. Despite the plethora of research focusing on Byzantine-resilient FL, the academic community has yet to establish a comprehensive benchmark suite, pivotal for impartial assessment and comparison of different techniques. This paper presents \ours, a scalable, extensible, and easily configurable benchmark suite that supports researchers and developers in efficiently implementing and validating novel strategies against baseline algorithms in Byzantine-resilient FL. \ours contains built-in implementations of representative attack and defense strategies and offers a user-friendly interface that seamlessly integrates new ideas. Using \ours, we re-evaluate representative attacks and defenses on wide-ranging experimental configurations (approximately 1,500 trials in total). Through our extensive experiments, we gained new insights into FL robustness and highlighted previously overlooked limitations due to the absence of thorough evaluations and comparisons of baselines under various attack settings.
We maintain the source code and documents at \url{https://github.com/lishenghui/blades}.
\end{abstract}


\input{contents}

\bibliographystyle{IEEEtran}
\bibliography{references}
	
\end{document}

%% file: includes.tex
\usepackage{nicefrac}
\usepackage{microtype}
\usepackage[tableposition=top]{caption}
\usepackage{subcaption}
\usepackage{comment}
\usepackage[switch]{lineno}
\usepackage{thmtools, thm-restate}
\usepackage{times}
\usepackage{wrapfig,lipsum,booktabs}
\usepackage{amsmath,bm}
\usepackage{xspace}
\usepackage{adjustbox}
\usepackage{makecell}
\usepackage{colortbl}
\usepackage{multirow}
\usepackage{rotating}
\usepackage{enumitem}
\usepackage{bbding}
\usepackage{siunitx}
\usepackage{colortbl}
\usepackage{arydshln}
\usepackage{fancyhdr}
\usepackage{tabularx}
\usepackage{array}
\usepackage{listings}
\usepackage[hidelinks]{hyperref}
\usepackage{url}
\usepackage[utf8]{inputenc}
\usepackage[T1]{fontenc}
\usepackage[outline]{contour}
\usepackage{soul}
\usepackage{algorithm}
\usepackage{algpseudocode}
\usepackage[flushleft]{threeparttable} 
\usepackage{tcolorbox}

\newtcolorbox{mybox}{colback=black!5!white,colframe=black,bottomrule=.25mm,toprule=.25mm,leftrule=.25mm,rightrule=.25mm,left=.25mm,right=.25mm,top=-.25mm,bottom=-.25mm}

\newcommand{\myparatight}[1]{\smallskip\noindent{\bf {#1}:}~}

\hypersetup{colorlinks,urlcolor=blue,citecolor=blue,breaklinks=true}

\newcolumntype{H}{>{\setbox0=\hbox\bgroup}c<{\egroup}@{}}
\newcommand{\noaistats}[1]{}  %

\definecolor{darkgreen}{rgb}{0,0.4,0.0}
\definecolor{darkblue}{rgb}{0,0.1,0.3}
\definecolor{darkred}{rgb}{0.7,0.0,0.0}

\definecolor{codegreen}{rgb}{0,0.6,0}
\definecolor{codegray}{rgb}{0.5,0.5,0.5}
\definecolor{codepurple}{rgb}{0.58,0,0.82}
\definecolor{backcolour}{rgb}{0.95,0.95,0.92}

\definecolor{deepblue}{rgb}{0,0,0.5}
\definecolor{deepred}{rgb}{0.6,0,0}
\definecolor{deepgreen}{rgb}{0,0.5,0}

\DeclareFixedFont{\ttb}{T1}{txtt}{bx}{n}{8} 
\DeclareFixedFont{\ttm}{T1}{txtt}{m}{n}{8}  
\lstdefinestyle{mystyle}{
  backgroundcolor=\color{backcolour},   
  commentstyle=\color{deepgreen},
  keywordstyle=\color{magenta},
  numberstyle=\footnotesize\color{codegray},
  stringstyle=\color{codepurple},
  emph={AdversaryCallback ,ClientCallback},          
  emphstyle=\ttb\color{deepred},    
  basicstyle=\ttm,
  morekeywords={self},              
  keywordstyle=\ttb\color{deepblue},
  breakatwhitespace=false,
  breaklines=true,
  captionpos=b,
  keepspaces=true,
  numbers=left,
  frame=single,
  numbersep=5pt,
  showspaces=false,
  showstringspaces=false,
  showtabs=false,
  xleftmargin=0.5cm,
  tabsize=2
}

\lstdefinelanguage{yaml}{
    basicstyle=\ttm,
    breaklines=true,
    frame=single,
    morestring=[b]',
    morestring=[b]",
    morecomment=[l]\#,
    morekeywords={grid_search},
    keywordstyle=\color{deepred},
    stringstyle=\color{red},
    commentstyle=\color{deepgreen},
    tabsize=4,
    showstringspaces=false,
}

\newcommand{\paragraphb}[1]{\noindent{\bf #1} }

\newcommand{\loss}{\ell}

\newcommand{ \algfont}[1]{\texttt{\hyphenchar \font=\defaulthyphenchar #1}}
\newcommand{\aggfont}[1]{\texttt{#1}}

\DeclareMathAlphabet{\mathpzc}{OT1}{pzc}{m}{it}

\newcommand{\bw}{\bm{w}}

\newcommand{\update}{\bm{\Delta}}

\newcommand{\grad}{\triangledown}

\newcommand \expect {\mathbb{E}}
\newcommand{\fedavg}{\algfont{FedAvg}\xspace}
\newcommand{\fedsgd}{\algfont{FedSGD}\xspace}
\newcommand{\ours}{\aggfont{Blades}\xspace}

\newcommand{\gm}{\aggfont{GeoMed}\xspace}
\newcommand{\mean}{\aggfont{Mean}\xspace}
\newcommand{\median}{\aggfont{Median}\xspace}
\newcommand{\cc}{\aggfont{CC}\xspace}
\newcommand{\bucketing}{\aggfont{Bucketing}\xspace}
\newcommand{\krum}{\aggfont{Krum}\xspace}
\newcommand{\tm}{\aggfont{TrimmedMean}\xspace}

\newcommand{\dnc}{\aggfont{DnC}\xspace}
\newcommand{\clippedclustering}{\aggfont{ClippedClustering}\xspace}
\newcommand{\signguard}{\aggfont{SignGuard}\xspace}
\newcommand{\code}[1]{{\texttt{\color[rgb]{0.03,0,1}{#1}}}}
\newcommand{\eg}{\emph{e.g.,}\xspace}
\newcommand{\ie}{\emph{i.e.,}\xspace}

%% file: contents.tex
\section{Introduction}

Federated learning (FL)~\cite{mcmahan2017communication,konevcny2015federated} has emerged as a compelling paradigm, allowing for collaborative machine learning model construction by leveraging distributed data across a diverse range of client devices, from IoT and edge devices to mobile phones and computers. 
FL enables on-device machine learning without the need to migrate
local data to a central cloud server, which is suitable for distributed IoT devices collecting massive sensor data. FL for IoT significantly impacts many existing and future IoT applications, including smart grids, smart transportation, smart health, and augmented reality ~\cite{khan2021federated,li2021byzantine}. 
The FL process typically involves several iterative steps: Firstly, a central server distributes the current global model to the clients. Subsequently, the clients independently perform one or multiple local steps of stochastic gradient descent (SGD) using their local datasets and transmit the updates back to the server. The server then aggregates these local updates to generate a new global model, which serves as the basis for the next round of training. The FL paradigm allows clients to jointly train a machine learning model without disclosing their private data to the central server. Furthermore, FL exhibits improved communication efficiency compared to traditional distributed learning methods ~\cite{arjevani2015communication}, as it capitalizes on multiple local update steps before transmitting the updates~\cite{konevcny2016federated}. 

Due to the distributed characteristic of optimization, FL is vulnerable to Byzantine failures~\cite{lyu2020threats,rodriguez2023survey}, wherein certain participants may deviate from the prescribed update protocol and upload arbitrary parameters to the central server. This risk is notable in IoT applications~\cite{khan2021federated,li2021byzantine}, where their open architecture allows diverse interconnectivity, thereby potentially expanding the attack surface. In this context, typical FL algorithms like FedAvg~\cite{mcmahan2017communication}, which compute the sample mean of client updates for global model aggregation, can be significantly skewed by a single Byzantine client~\cite{li2023experimental}. The server thus requires Byzantine-resilient solutions to defend against malicious clients.

Depending on the adversarial goals, Byzantine attacks in FL can be classified into two categories: \textbf{targeted attacks} and \textbf{untargeted attacks}~\cite{jere2020taxonomy,lyu2020threats}. Targeted attacks, such as backdoor attacks, aim to manipulate the global model to generate attacker-desired misclassifications for some particular test samples~\cite{xie2019dba,bagdasaryan2020backdoor,andreina2021baffle}, while untargeted attacks aim to degrade the overall performance of the global model indiscriminately~\cite{fang2020local}. Our work focuses on \textit{untargeted attacks}, which is consistent with the majority of Byzantine-resilient research~\cite{blanchard2017machine,yin2018byzantine,chen2017distributed,li2021byzantine,pmlrv139karimireddy21a,sattler2020byzantine,9833647}. 
Henceforth, any reference to ``Byzantine'' will imply ``untargeted Byzantine''. 


In recent years, the field of FL has seen the emergence of various Byzantine-resilient approaches. They aim to protect distributed optimization from Byzantine clients and assure the performance of the learned models~\cite{rodriguez2023survey, hu2021challenges}. For instance, robust aggregation rules (AGRs) are widely used to estimate the global update from a collection of local updates while mitigating the impact of malicious behaviors. Typical AGRs include \gm~\cite{chen2017distributed}, \krum~\cite{blanchard2017machine}, \tm~\cite{yin2018byzantine}, and  \median~\cite{yin2018byzantine}. 
 Meanwhile, different attack strategies are emerging, striving to circumvent defense strategies~\cite {baruch2019little, xie2020fall}. For instance, the A Little Is Enough (ALIE) attack can bypass various AGRs by taking advantage of the empirical variance between clients' updates if such a variance is high enough, especially when the local datasets are not independent and identically distributed (non-IID)~\cite{baruch2019little,li2023experimental}. Thus, defending against adversarial attacks remains an open problem in FL~\cite{kairouz2021advances}.

Moreover, it has been shown that the experimental evaluation in existing studies may be insufficient to validate their robustness against diverse Byzantine attacks~\cite{khan2023pitfalls}, as they were only examined under specific experimental settings (\eg specific attack types, and hyper-parameter configurations). 
The study emphasizes that limited and narrowly focused evaluations might overlook certain vulnerabilities and provide an incomplete picture of their robustness against different threats. This underscores the pressing need for a unified benchmark suite that offers comprehensive assessments and fair comparison across various attacks and scenarios.


\textbf{Our work:} This paper presents \textit{\ours}, our open-source
benchmark suite to fill the identified gaps in existing experimental evaluations in Byzantine-resilient FL. Through \textit{\ours}, we conduct comprehensive experimental evaluations, scrutinizing a range of representative attacks and defense techniques. Specifically, we make the following two concrete contributions:

\myparatight{Contribution 1. \ours, a benchmark suite}We introduce \ours, an open-source benchmark suite for \textbf{B}yzantine-resilient federated \textbf{L}earning with \textbf{A}ttacks and \textbf{D}efenses \textbf{E}xperimental \textbf{S}imulation, which is specifically designed to fill the need for studying attack and defense problems in FL. \ours is built upon a versatile distributed framework, Ray, enabling effortless parallelization of single machine code across various settings, including single CPU, multi-core, multi-GPU, or multi-node, with minimal configuration requirements. This makes \ours efficient in terms of execution time, as client and server operations are executed in a parallel manner. In addition, \ours provides a wide range of attack and defense mechanisms and allows end users to plug in customized or new techniques easily. We illustrate the user-friendly nature of \ours through examples and validate its scalability with respect to clients and computational resources. The results highlight that \ours can effectively handle large client populations and computational resources.

\myparatight{Contribution 2. Comparative case studies} Using \ours, we conduct an exhaustive re-evaluation of six representative AGRs, encompassing both three classical and three contemporary methods, against six attacks on three datasets. Additionally, we also inspect key factors and risks that might affect the robustness of AGRs, including data heterogeneity, differential privacy (DP) noise, momentum, and the risk of gradient explosion. The results unveil new insights. The \textbf{key findings} from our experiments can be summarized as follows:

\begin{enumerate}
    \item The effectiveness of adversarial attacks depends on several factors,
    including the choice of dataset, defensive countermeasure, and the specific FL algorithm employed.
    \item The robustness of defenses in existing studies may be overrated owing to their insufficiency in comprehensive evaluation under wide-ranging settings.
    \item Additionally, various factors, including data heterogeneity, differential privacy (DP) noise, and momentum, exert considerable influence on the Byzantine resilience of defense strategies.
\end{enumerate}

The key findings underscore the intrinsic complexities and nuances in ensuring Byzantine resilience in FL and further emphasize the importance of a unified benchmark like \ours that enables comprehensive evaluations on attack and defense techniques.

\input{background}

\input{blades}

\input{experiments}

\subsection{Scalability Evaluation}

We evaluate the scalability of \ours along two lines: First, we evaluate the training time for a global round with an increasing number of clients and computing resources.
Second, we evaluate the training time per round with an increasing number of GPUs.

To assess the scalability with respect to the number of clients, we conduct simulations with a fixed set of resources and vary the number of clients. As shown in Table~\ref{scal_client}, the number of clients increases from 16 to 512, which greatly outnumbers the available CPUs and GPUs. The results show that with a linear increase in the number of clients, the average training time of each global round increases linearly with little standard deviation, indicating the stable and efficient communication and task distribution implementation in \ours.

\begin{table}[ht]
\caption{\small Experiment for client scalability: average and standard deviation of training time per global round with increasing numbers of clients.}
\vspace*{-.4cm}
\begin{center}
\setlength\extrarowheight{2pt}
\begin{adjustbox}{width=.9\linewidth,center}
\begin{tabular}{c|ccccccc}
\hline
    \textbf{\# Clients} & 16 & 32 & 64 & 128 & 256 & 512 \\
\hline
\hline
    Avg (seconds) & 1.41 & 2.48 & 4.51 & 9.10 & 17.58 & 34.53\\
    Std (seconds) & 0.05 & 0.07 & 0.11 & 0.18 & 0.38 & 0.73\\
\hline
\end{tabular}
\end{adjustbox}
\label{scal_client}
\end{center}
\end{table}

\begin{figure}[t]
\centering
\includegraphics[width=.85\linewidth]{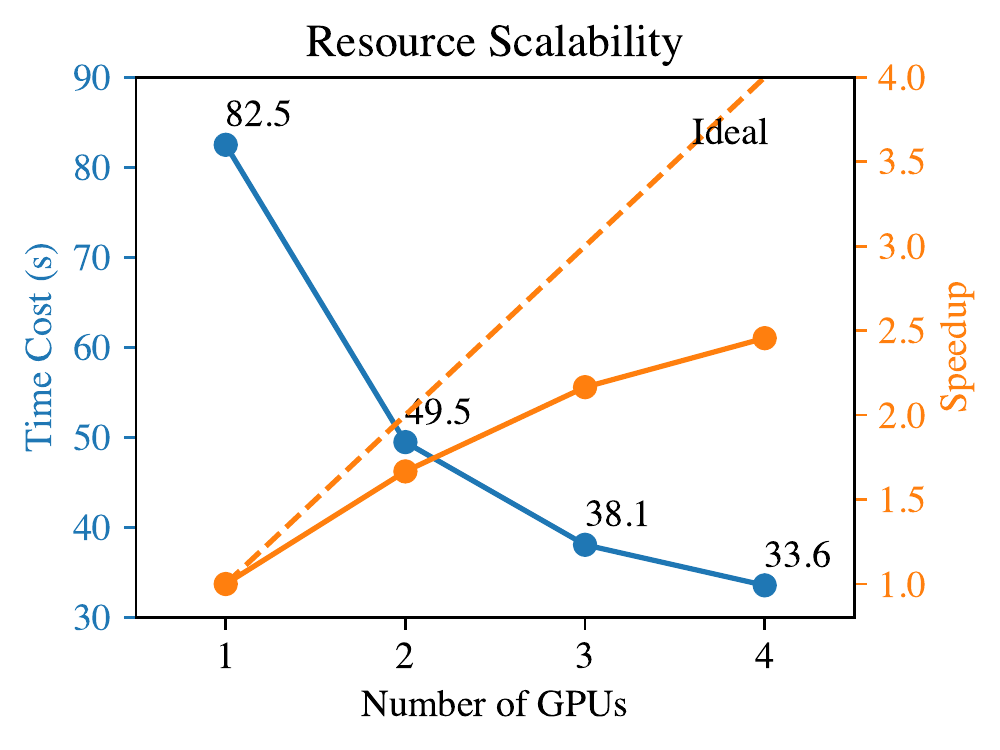}
\vspace*{-.2cm}
\caption{Resource scalability of \ours. Blue: Average time cost per global round decreases with the number of GPUs. Orange: The speedup increases with the number of GPUs.}
\label{time_gpu}
\end{figure}


To evaluate the resource scalability of \ours, we design a simulation task involving 480 clients to be executed on GPUs. Fig. \ref{time_gpu} shows the time cost of each global round with different numbers of GPUs and the associated speedups. The time cost for each global round reduces from 82.5s to 49.5s when the number of GPUs increases from 1 to 2, and the time cost reduces more if more GPUs are added. The speedup achieved does not align precisely with the number of GPUs utilized, primarily because FL is not an entirely parallelizable algorithm. Consequently, a substantial portion of serial work, such as the communication of local updates and model aggregation, hinders the potential speedup. Nonetheless, the increase in computing resources significantly reduces the time cost of the simulation, and \ours is scalable with computing resources.

\input{related_work}

\section{Conclusion and Future Work}

In this paper, we introduced an open-source benchmark suite, namely \ours, to address the research gap concerning attack and defense problems in FL. The \ours framework offers usability, extensibility, and scalability that facilitate the implementation and expansion of novel attack and defense techniques. By utilizing \ours, we conducted a comprehensive evaluation of prominent defense techniques against state-of-the-art attacks, yielding insightful findings regarding the resilience of Byzantine-resilient FL.
Furthermore, we showed that \ours is scalable in terms of both the number of clients and computing resources. In the future, we will continue our efforts to address more security threats with our new releases of \ours, and integrate more cutting-edge methods for comprehensive benchmarking.





%% file: background.tex
\section{Background and Related Work}

\subsection{FL and Optimization}
	\label{sec_formulation}

In FL, a collection of clients collaboratively learn a shared global model by leveraging their private datasets in a distributed manner, assisted by the coordination of a central server. The goal is to find a parameter vector $\bm{w}$ that minimizes the following distributed optimization model:
	\begin{equation}
		\label{obj}
		\min\limits_{\bm{w}} F(\bm{w}) :=  \frac{1}{K} \sum_{k \in [K]}  F_k(\bm{w}),
	\end{equation}
	where $K$ represents the total number of clients and $F_k(\bm{w}) = \expect_{\bm{z}\sim \mathcal{D}_k} [\loss(\bw; \bm{z})]$ denotes the expected risk of the $k$-th client. Here,
    $\mathcal{D}_k$ is the data distribution for the $k$-th client 
	and $\loss(\cdot;\cdot)$ is a user-specified loss function.
\begin{algorithm}[t]
    \algdef{SE}[SUBALG]{Indent}{EndIndent}{}{\algorithmicend\ }%
    \algtext*{Indent}
    \algtext*{EndIndent}
    \setlength{\abovedisplayskip}{0pt}
    \setlength{\belowdisplayskip}{0pt}
    \setlength{\abovedisplayshortskip}{0pt}
    \setlength{\belowdisplayshortskip}{0pt}
    \caption{A \fedavg-family Algorithm for FL}
    \begin{flushleft}
    \hspace*{\algorithmicindent} \textbf{Input:} $K, T$, $\bw^{0}$, CLIENTOPT, SERVEROPT
    \end{flushleft}
    \label{alg}
    \begin{algorithmic}[1]
        \For{each global round $t \in [T]$}
        \State{Select a subset $S_t$ from $K$ clients at random}
        \For{each client $k \in S_t$ \textbf{in parallel}}
        \State{$\bm{w}_k^t \leftarrow \bm{w}^t$}
        \For{$E_l$ local rounds} \label{ag_g_data}
        \State{Compute an estimate $g_k( \bm{w}_{k} ^ t)$ of $\grad F_k( \bm{w}_{k} ^ t)$}
        \State{$ {\displaystyle \bm{w}_{k} ^ t \leftarrow  \text{CLIENTOPT} (\bm{w}_{k} ^ t, g_k( \bm{w}_{k} ^ t), \eta_l, t)}$}
        \EndFor
        \State $\update_k^t \leftarrow \bm{w}_{k}^t - \bm{w}^t$ \label{update_cal}

        \State{Send $\Delta_k^t$ back to the server} \label{ag_g_update}
        \EndFor
        \State $\Delta^{t+1} \leftarrow \text{AGG}(\{\Delta_k^t\}_{k \in S_t})$

        \State $\bm{w}^{t+1} \leftarrow \text{SERVEROPT}(\bm{w}^{t}, -\Delta^{t+1}, \eta_g, t)$
        \EndFor
        \State \Return $\bw^{T}$
    \end{algorithmic}
    \label{opt}
\end{algorithm}

The most popular algorithms in the literature that solve \eqref{obj} are the \fedavg-family algorithms~\cite{mcmahan2017communication,reddi2020adaptive,ju2023accelerating}. As shown in Algorithm~\ref{alg}, at $t$-th round of communication, a subset of clients $S_t$ is selected, typically through a random sampling process. The server then broadcasts its current global model parameters $\bm{w}^t$ to each selected client. Simultaneously, the clients independently perform local optimization on their respective private data, aiming to minimize their own empirical loss. This process involves multiple local rounds, denoted as $E_l$, where the clients compute an estimate $g_k(\bm{w}_k^t)$ of the gradient $\nabla F_k(\bm{w}_k^t)$ from their local data. The client model's $\bm{w}_k^t$ are iteratively updated based on the estimated gradient and a client-specific learning rate $\eta_l$. The computed local model updates, denoted as $\update_k^t$, are then transmitted back to the server. The server aggregates these updates using an aggregation rule, often averaging aggregation~\cite{mcmahan2017communication}, to generate a global update. This update represents a direction for the global optimizer, capturing the collective knowledge of the participating clients. Subsequently, the server employs the global optimizer, denoted as $\text{SERVEROPT}$, to update the global model's parameters $\bm{w}^t$ using the negative of the aggregated updates, denoted by $-\update_k^{t+1}$ (which is called ``pseudo-gradient''~\cite{reddi2020adaptive}), and a global learning rate $\eta_g$. By iterating this process for multiple rounds, the \fedavg-family algorithms refine the global model by leveraging the clients' distributed computing capabilities and decentralized datasets.

Our study adopts the full client participation paradigm in alignment with previous research~\cite{liconvergence}. As such, every client is actively engaged in each round of local training, ensuring that $|S_t| = K$ as per Algorithm~\ref{alg}. The rationale behind this choice is grounded in a prevailing assumption of Byzantine-resilient studies in FL, i.e., the number of malicious updates for aggregation is less than half during each round. Selecting subsets at random risks contravening this foundational assumption, given the inherent possibility of inadvertently favoring an excessive proportion of adversarial clients over their benign counterparts~\cite{li2023experimental}. 




\subsection{Scope of our work: Byzantine Attacks and Defenses in FL}
\label{scope}
Byzantine attacks pose a significant threat to FL due to its distributed optimization nature~\cite{lyu2020threats,rodriguez2023survey}.  
In general, the malicious clients may upload arbitrary parameters to the server to degrade the global model's performance. Hence, in Algorithm~\ref{alg}, the \fedavg-family algorithm, Line~\ref{update_cal} can be replaced by the following update rule:
	\begin{align}
		\label{inner_solution}
		\update_k^t &\leftarrow
		\begin{cases}
			\star            & \text{if $k$-th client is Byzantine,} \\
			\bm{w}_{k}^t - \bm{w}^t & \text{otherwise,}
		\end{cases}
	\end{align} where $\star$ represents arbitrary values.



As aforementioned, the scope of this work is on untargeted Byzantine attacks, where the adversary's objective is to minimize the accuracy of the global model for any test input~\cite{baruch2019little,fang2020local,shejwalkar2021manipulating,li2023experimental}. Various attack strategies have been proposed to explore the security vulnerabilities of FL, taking into account different levels of the adversary's capabilities and knowledge~\cite{pmlrv139karimireddy21a,li2019rsa,baruch2019little,shejwalkar2021manipulating}. For instance, with limited capabilities and knowledge and without having access to the training pipeline, the adversary can manipulate a single client's input and output data. In more sophisticated attacks, the adversary possesses complete knowledge of the learning system and designs attack strategies to circumvent defenses. Below, we detail here some typical attacks:

\myparatight{LabelFlipping~\cite{pmlrv139karimireddy21a}} The adversary simply flips the label of each training sample~\cite{fang2020local}. Specifically, a label $l$ is flipped as $L-l-1$, where $L$ is the number of classes in the classification problem and $l=0,1,..., L-1$.

\myparatight{SignFlipping~\cite{li2019rsa}} The adversary strives to maximize the loss via gradient ascent instead of gradient descent. Specifically, it flips the gradient's sign during the local updating step. 

\myparatight{Noise~\cite{li2021byzantine}} The adversary samples some random noise from a distribution (e.g., Gaussian distribution) and uploads it as local updates.

\myparatight{ALIE~\cite{baruch2019little}} The adversary takes advantage of the empirical variance among benign updates and uploads a noise within a range without being detected. For each coordinate $i \in [d]$, the attackers calculate mean ($\mu_i$) and std ($\delta_i$) over benign updates and set malicious updates to values in the range $(\mu_i - z^{max}\delta_i, \mu_i + z^{max}\delta_i)$, where $z^{max}$ ranges from 0 to 1, and is typically obtained from the Cumulative Standard Normal Function. The $i$-th malicious update is then obtained by $\Delta_{k,i}^t \leftarrow \mu_i - z^{max} \mu_i$.

\myparatight{IPM~\cite{xie2020fall}}
The adversary seeks the negative inner product between the true mean of the updates and the output of the aggregation rules so that the loss will at least not descend. Assuming that the attackers know the mean of benign updates, a specific way to perform an IPM attack is
	\begin{equation}
		\update_1^t = \dots = \update_M^t = - \frac{\epsilon}{K - M}\sum_{i = M+1}^K \update_i^t,
	\end{equation} assuming that the first $M$ clients are malicious, $\epsilon$ is a positive coefficient controlling the magnitude of malicious updates.

\myparatight{MinMax~\cite{shejwalkar2021manipulating}} Similar to ALIE, the adversary strives to ensure that the malicious updates lie close to the clique of the benign updates. The difference is that MinMax re-scales $z^{max}$ such that the maximum distance from malicious updates to any benign updates is upper-bounded by the maximum distance between any two benign updates.

As for defenses, robust aggregation rules (AGRs) are widely applied to make a Byzantine-resilient estimation of the true updates and exclude the influence of malicious updates~\cite{blanchard2017machine,yin2018byzantine,chen2017distributed,li2021byzantine,sattler2020byzantine}. While other research directions, such as trust-based strategies~\cite{cao2021fltrust,9887909,sageflow} and variance-reduced algorithms~\cite{gorbunov2023variance,9153949}, are worth exploring, our study primarily focuses on AGRs. This is because most existing studies predominantly consider AGR-based defenses, and we specifically examine \median~\cite{yin2018byzantine}, \tm~\cite{yin2018byzantine}, \gm~\cite{pillutla2022robust}, \dnc~\cite{shejwalkar2021manipulating}, \clippedclustering~\cite{li2023experimental}, and \signguard~\cite{xu2022byzantine} in this work.

 
\subsection{Unifying Byzantine-resilient FL with Traditional DL }

The study of Byzantine-resilient FL can be traced back to traditional distributed learning (DL), where a central server distributes data to workers who perform gradient estimation; the gradients are then aggregated by the server for model update~\cite{ alistarh2018byzantine,yin2018byzantine,chen2017distributed}. FL originally emerged as an extension of distributed learning to address the limitations imposed by communication constraints and privacy concerns associated with decentralized data ownership~\cite{yin2021comprehensive}. Although FL and traditional distributed learning are employed in different application domains~\cite{li2020review}, they share similar security vulnerabilities stemming from Byzantine attacks due to the distributed nature of optimization. Furthermore, many existing techniques initially proposed for studying Byzantine-resilient distributed learning~\cite{blanchard2017machine,chen2017distributed,yin2018byzantine,bernstein2018signsgd,baruch2019little} have now found extensive application in the defense mechanisms utilized in FL~\cite{pillutla2022robust,fang2020local,shejwalkar2021manipulating,9833647,li2023experimental}. Therefore, it is important to examine FL and traditional distributed machine learning together when it comes to Byzantine resilience.

Benefiting from the generality of Algorithm~\ref{alg}, obtaining traditional distributed learning algorithms is straightforward. For example, by assuming both ``CLIENTOPT'' and ``SERVEROPT'' as a gradient descent step and setting $E_l=1$ and $\eta_l=1$, Algorithm~\ref{alg} simplifies to the naive distributed SGD with gradient aggregation~\cite{chen2017distributed}. In contrast, setting $\eta_g=1$ leads to the naive \fedavg~algorithm. This connection enables the generalization of traditional techniques, such as robust aggregation, from traditional distributed learning to suit the requirements of FL.

    
\subsection{Benchmarks for Byzantine-resilient FL}

\input{tables/benchmark_table}

    Recently, various FL benchmarks have been introduced with different emphases and scopes including scalability (\eg Fedscale~\cite{lai2022fedscale}), heterogeneity (\eg B-FHTL~\cite{yao2022benchmark} and pFL-bench~\cite{chen2022pfl}), privacy (\eg PrivacyFL~\cite{mugunthan2020privacyfl} and Pysyft~\cite{ziller2021pysyft}), and security (\eg AggregaThor~\cite{damaskinos2019aggregathor}, FedMLSecurity~\cite{han2023fedmlsecurity}, and Backdoor Bench~\cite{qin2023revisiting}). Concerning Byzantine-resilience, AggregaThor~\cite{damaskinos2019aggregathor}, FedMLSecurity~\cite{han2023fedmlsecurity} are the most relevant benchmarks to our work. However, they both fall short across core dimensions (Table~\ref{benchmark_table}). AggregaThor is mainly limited in the versatility of attacks/defenses customization and FL algorithms, as it only supports traditional distributed gradient aggregation in \fedsgd. In contrast, FedMLSecurity predominantly centers on the \fedavg-family optimizers but overlooks gradient aggregation techniques. Although FedMLSecurity is compatible with various FL optimizers (\eg FedPROX~\cite{li2020federated} and FedNOVA~\cite{wang2020tackling}), we claim that \fedsgd and \fedavg~represent the foundational algorithms for benchmarking Byzantine-resilient FL, given the substantial volume of attacks and defenses constructed around them. Furthermore, AggregaThor and FedMLSecurity lack user-friendly mechanisms for hyperparameter tuning, resulting in great engineering efforts when benchmarking across varied configurations. In response to these gaps, we have designed and developed a novel benchmark suite, \ours, that addresses the aforementioned limitations and offers a more comprehensive evaluation for Byzantine-resilient FL.

%% file: tables/benchmark_table.tex
\begin{table}[!t]
\newcommand \alignline{}
\newcommand \thickvline{ \color{black}\vline width 1pt}
\newcommand{\rotbox}[1]{\rotatebox[origin=c]{0}{#1}}
\caption{Comparing \ours with existing benchmarks for Byzantine-resilient FL}
\setlength\extrarowheight{3pt}
\setlength{\arrayrulewidth}{0.3mm}
\begin{adjustbox}{width=0.98\linewidth,center}
    \centering
    \begin{tabular}{c|cc|c} 
        \hline
         \textbf{Features} &  \textbf{AggregaThor}~\cite{damaskinos2019aggregathor}& \textbf{FedMLSecurity}~\cite{han2023fedmlsecurity} & \textbf{\ours}  \\ 
        \hline
        \textbf{Year}  &  2019&  2023 & 2023 \\ 
        \textbf{\makecell{ML Backend}} &  TensorFlow &  Pytorch & Pytorch\\ 
        \textbf{\makecell{Distributed  Backend}} &  MPI &  MPI & Ray \\  
        \textbf{\makecell{Flexible APIs}} &   \XSolidBrush &  \Checkmark & \Checkmark \\ 
        \textbf{\makecell{\fedsgd Algorithm}} & \Checkmark  & \XSolidBrush  & \Checkmark \\ 
        \textbf{\makecell{\fedavg-family Algorithms}} &  \XSolidBrush &  \Checkmark & \Checkmark \\ 
        \textbf{\makecell{Hyperparameter Tuning}} &  \XSolidBrush &  \XSolidBrush & \Checkmark \\ 
         \hline
    \end{tabular}
    \label{benchmark_table}
\end{adjustbox}

\end{table}

%% file: blades.tex
\begin{figure}[!t]
	\centering
	\includegraphics[width=.99\linewidth]{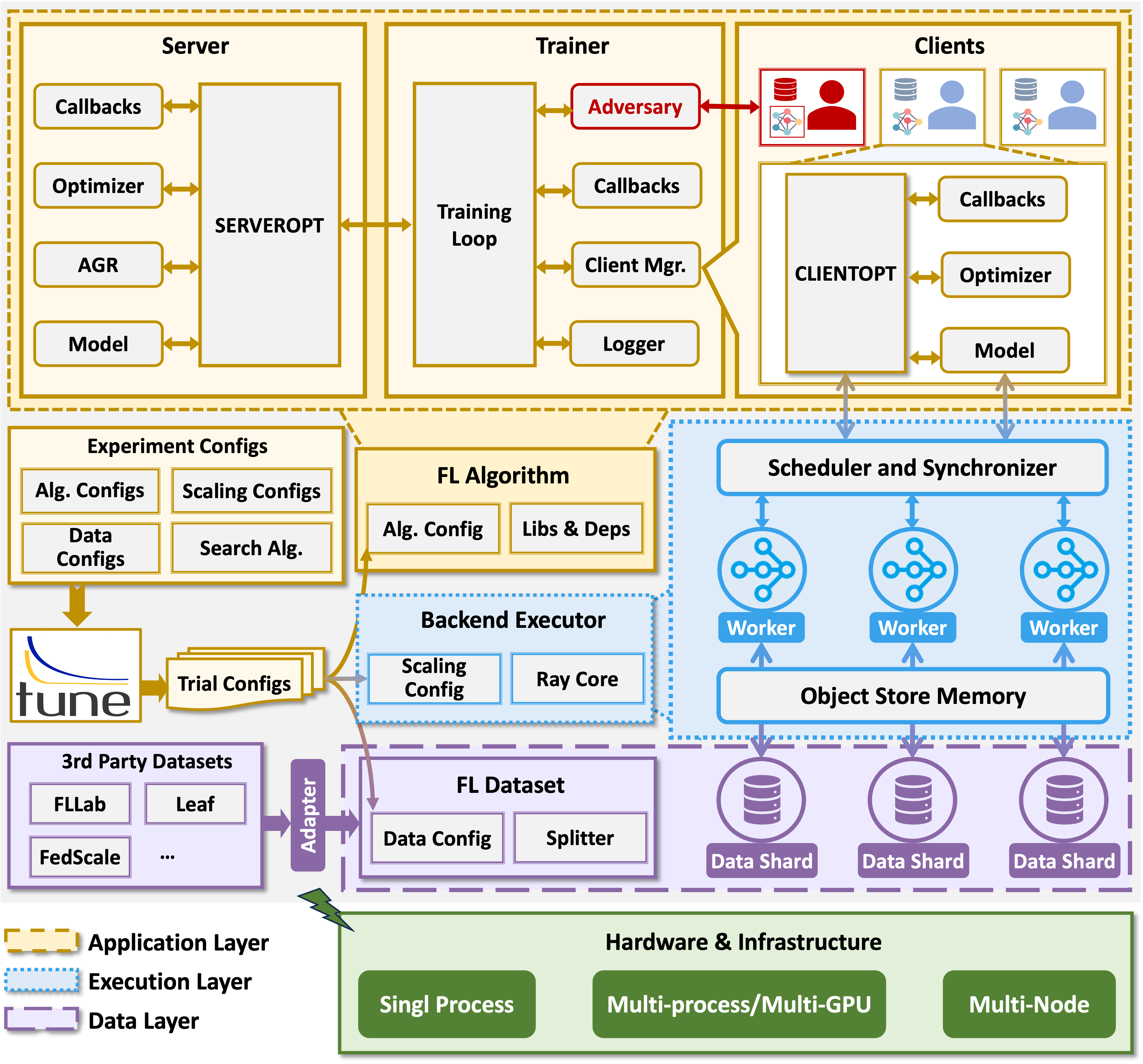}
	\caption{Overview of the \ours architecture comprising the Application, Execution, and Data layers. 
	}
	\label{arch}
\end{figure}

\section{The Design of \ours}
\label{sec_design}
In this section, we first outline our design goals for \ours. Following that, we delve into the structure of Blades, elucidating its layered architecture and key components. Finally, we demonstrate the implementation of both attacks and defenses with \ours.

\subsection{Design Goals}

We designed and implemented \ours with the following goals in mind:


\myparatight{Usability} The benchmark suite should enable researchers to effortlessly set up experiments, search, and tune hyperparameters. To enable efficient comparisons, it should include implementations of the most representative algorithms (e.g., \fedsgd and \fedavg) with attack and defense strategies.

\myparatight{Extensibility} As new attack and defense strategies are discovered, it should be relatively straightforward to incorporate them into the suite. Additionally, the design should readily accommodate the integration of new datasets, models, and FL algorithms.


\myparatight{Scalability} The suite should exhibit scalability in terms of both clients and computing resources. 
Scalability with clients refers to the ability to accommodate a large and diverse set of clients participating in the learning process. Scalability with computing resources entails efficiently utilizing and adapting to different hardware setups.


\subsection{Core Framework Architecture}

In pursuit of our design goals, particularly extensibility and scalability, we structured the system into three distinct layers: the Application layer, Execution layer, and Data layer. An overview of the architecture of \ours is illustrated in Fig.~\ref{arch}. This tripartite architecture is rooted in our intention to delineate and modularize specific objectives. The Application layer emphasizes extensibility, allowing for straightforward configuration and integration of a range of FL-related functionalities and features. The Execution layer is tasked with ensuring system scalability, adeptly accommodating extensive workloads. Finally, the Data layer streamlines data ingestion and preprocessing in the distributed setting.

\subsubsection{Application Layer}
It is the top layer of \ours and provides a user-friendly interface for designing and deploying FL algorithms. The main abstractions in this layer include:

\myparatight{Server} A server is an object that aggregates model updates from multiple clients and performs global optimization. Once a local training round is finished, it gathers the model updates and takes one iteration step.  Defense strategies, such as AGRs, are usually applied here to eliminate the impact of malicious updates.

\myparatight{Trainer} The trainer encapsulates the optimization process for a particular FL algorithm. A trainer manages key aspects of the training loop, including interactions with the server, client manager, and state synchronization between the server and clients. Each trainer corresponds to a specific FL algorithm, such as \fedavg, and can be configured with various hyperparameters to control the local training process. It also allows customization with callbacks invoked at specific points during training, such as after each local training round or server optimization step. The ``Adversary'' component in the Trainer can control a subset of clients to perform malicious operations.

\myparatight{Client} The client acts as a participant in the FL process. We provide the client-oriented programming design pattern~\cite{he2020fedml} 
to program the local optimization of clients during their involvement in training or coordination within the FL algorithm. This pattern allows end users to specify and execute certain types of attacks easily. Additionally, the interface we offer allows users to tailor the behaviors of Byzantine clients.

\begin{figure}[!t]
\begin{subfigure}[b]{\linewidth}
  \begin{lstlisting}[
language=yaml,
linewidth=0.99\linewidth,
]
stop: training_round: 2000 # Communication rounds
config:    
    global_model: ResNet10
    num_malicious_clients: grid_search: [0, 5]
    # Configuring AGR and optimizer for Server 
    server_config:   
      AGR: grid_search: [Mean, Median]
      optimizer:           # SERVEROPT
        type: SGD
        lr_schedule: [[0, 0.1], [1500, 0.1]]
    # Specify adversarial attack and parameters
    adversary_config:
      grid_search:
        - type: LabelFlipAdversary
        - type: IPMAdversary
            alpha: grid_search: [0.1, 100]
\end{lstlisting}
    \end{subfigure}
\caption{An example YAML configuration snippet for simulating LabelFlipping and IPM attacks with various hyperparameter settings. \ours is fully configurable and allows grid search for hyperparameter settings and experiment specifications.}
\label{config_code}
\end{figure}

The application layer has several dependencies that provide a variety of functionalities. Particularly, we adopt the Tune library\footnote{\url{https://docs.ray.io/en/latest/tune}}~\cite{liaw2018tune} for experiment configuration and hyperparameter tuning at any scale. As an example, Fig.~\ref{config_code} shows a configuration file for simulating the LabelFlipping~\cite{li2023experimental} and IPM~\cite{baruch2019little} attacks with different hyperparameter settings. With the help of Tune, \ours reads the file and parses the configurations to generate a series of experimental trials. The trials are then scheduled to execute on the execution layer. This functionality facilitates the creation of multiple combinations of hyperparameters and configurations found within the grid. Specifically, in the given example, the grid searching areas of configuration include \code{num\_malicious\_clients}'', \code{AGR}'', ``\code{adversary\_config}'', and  ``\code{alpha}''. All considered, this culminates in a total of 12 trials generated simultaneously.


\subsubsection{Execution Layer} 

\input{execution_layer}

\subsubsection{Data Layer} The data layer facilitates data pre-processing and loading for distributed training with the execution backend. It supports both IID and non-IID splitting for studying homogeneous and heterogeneous scenarios, respectively. At the beginning of the training, the dataset is separated into multiple shards and pre-allocated to workers' memory to allow fast data loading. In addition, we provide an adapter to import datasets from well-known 3rd-party benchmarks in FL, such as Leaf~\cite{caldas2018leaf}, FedScale~\cite{lai2022fedscale}, and FedLab~\cite{JMLR:v24:22-0440}.

\subsection{Implementation of Attacks and Defenses}

\begin{figure}[t]
	\centering
	\includegraphics[width=\linewidth]{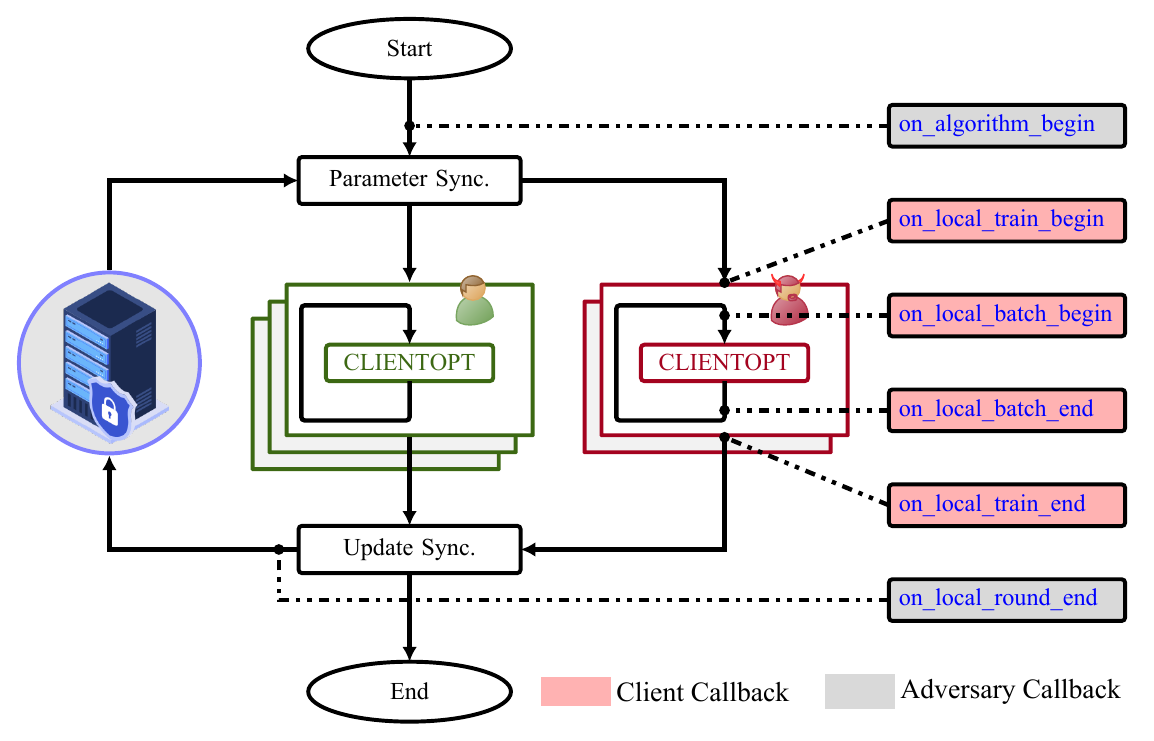}
    \vspace*{-.5cm}
	\caption{The pipeline for implementing attacks in \ours. We define some time points when
users can register executable methods to perform customized attacks. At specific time points, client callbacks and adversary callbacks are triggered to invoke the registered methods.}
	\label{pipeline}
\end{figure}


\subsubsection{Implementing Attacks}
We note that adversarial attackers may perform some self-defined manipulation before or after specific time points. For instance, LabelFlipping~\cite{pmlrv139karimireddy21a} attacks are typically executed at the beginning of batch forward propagation, while SignFlipping~\cite{li2019rsa} attacks are carried out immediately after backpropagation. To address this, we have designed a unified pipeline integrated with a callback mechanism, which enables actions to be performed at various stages of training, as depicted in Fig.~\ref{pipeline}. This design offers extensibility to facilitate customization, where the minimal pipeline focuses on repetitive local training and server-side optimization while the malicious behaviors are defined through the callback mechanism. Users only need to override specific callback methods to execute a custom attack without modifying the pre-defined logic.

\ours incorporates a dual-tiered callback mechanism: at the client and adversary levels. Elementary attacks (\eg LabelFlipping and SignFlipping), operate solely on local data or model parameters. Therefore, their implementation is streamlined by registering specific behaviors to client objects, enabling concurrent execution across multiple workers. As an illustrative instance, the upper panel of Fig.~\ref{atk_sim_code} shows a code snippet that exemplifies the implementation of a LabelFlipping attack on a classification task encompassing 10 distinct classes. Through a straightforward customization of the "\code{on\_batch\_begin()}" callback method, users can easily modify the labels from class $i$ to $9-i$ during local training. 

For more sophisticated attacks such as colluding attacks, clients in the distributed environment need to exchange data, coordinate actions, and synchronize their activities to make decisions regarding malicious actions. A straightforward solution to simulate such attacks is allowing inter-client communication using the remote function mechanism in Ray. However, this will make the simulation more complex and limit the scalability of the system. One possible consequence is the occurrence of deadlocks~\cite{holt1972some}, i.e., when two or more clients are waiting for each other to release a resource or respond to a communication, none of them can proceed. 

\begin{figure}[t]
\begin{subfigure}[b]{\linewidth}
  \begin{lstlisting}[
    language=Python,
    linewidth=0.99\linewidth,
    ]
from blades.clients import ClientCallback
class LabelFlipCallback(ClientCallback):
    def on_batch_begin(self, data, target):
        # Return input data with flipped labels for the current batch (assuming 10 labels).
        return data, 9 - target
\end{lstlisting}
\vspace{-2pt}
\end{subfigure}

\begin{subfigure}[b]{\linewidth}
  \begin{lstlisting}[
language=Python,
linewidth=0.99\linewidth,
]
from blades.adversary import AdversaryCallback
class ALIECallback(AdversaryCallback):
    def on_local_round_end(self, trainer):
        # Compute the dimensional mean and std over client updates, then generate malicious updates by adding std to the mean.
        updates = trainer.get_updates()
        mean = updates.mean(dim=0)
        std = updates.std(dim=0)
        updates = mean + std
        trainer.save_malicious_updates(updates)
\end{lstlisting}
    \end{subfigure}
\vspace*{-.2cm}
\caption{Illustration of our callback mechanism used to simulate LabelFlipping (upper) and ALIE (lower) attacks. The design's flexibility enables easy customization by overriding methods associated with both client and adversary callbacks.
}
\label{atk_sim_code}
\end{figure}

Alternatively, we facilitate the implementation of sophisticated attacks by incorporating additional callbacks tailored for adversary entities. Distinguished from client callbacks, these adversary callbacks are executed within the driver program and allow convenient access to various system components and their states. As a result, this design simplifies the process of acquiring knowledge for high-level attacks. Fig.~\ref{pipeline} also shows two of the most essential adversary callbacks, specifically ``\code{on\_algorithm\_begin()}'' and ``\code{on\_local\_round\_end()}''. The former is triggered at the start of the algorithm and serves the purpose of initializing the adversary and setting up client callbacks. The latter is triggered upon the completion of a local round and allows for the modification of collected updates from malicious clients before proceeding to server-side optimization and defense operations. During the ``\code{on\_local\_round\_end()}'' callback, one can potentially access honest updates and other system states in a read-only manner to launch omniscient attacks. The lower panel of Fig.~\ref{atk_sim_code} shows an example of implementing the ALIE~\cite{baruch2019little} attack using this adversary callback. 


\subsubsection{Implementing Defenses}
Defenses in the context of our study stem from multiple facets, posing challenges to the establishment of a standardized pipeline akin to the one employed for attacks. Nevertheless, certain indispensable steps are involved in this process, namely update aggregation and global model optimization, although the specific methodology for each step may vary. The update aggregation step combines the locally collected updates, while the global model optimization step performs an optimization procedure based on the aggregated result.

 As such, \ours introduces a foundational abstraction of the server entity, encompassing essential components including global model, AGR, and optimizer. This architecture permits the extension of functionalities through the utilization of sub-classing, thereby facilitating the integration of advanced features.  It is noteworthy to mention that even in the minimal server implementation, we offer a configurable SGD optimizer and a variety of pre-defined AGRs. Furthermore, all the components are modularized and inheritable, allowing plug-and-play of different configurations. We believe our designs simplify the process of generating benchmark results with minimal effort.
 


%% file: execution_layer.tex
The execution backend is built upon a scalable framework Ray~\cite{moritz2018ray}
for training and resource allocation. Ray provides two key advantages for \ours: 1) It allows users to customize computing resources (e.g., CPUs and GPUs) to clients and servers conveniently; 2) It enables \ours to easily adapt to in-cluster large-scale distributed training, benefiting from the capabilities of the Ray cluster. The key components in the execution layer are:

\myparatight{Worker} A worker is a proxy that can be allocated with computing resources to execute the training pipeline for clients. Inherited from Ray Actor\footnote{\url{https://docs.ray.io/en/latest/ray-core/actors.html}}, the worker (essentially a Python process) is the smallest unit for resource management and distributed computing. In a distributed environment, multiple workers collaborate as a group. In case of a large number of clients and limited resources, multiple clients are mapped to one worker, and their local training tasks will be scheduled to run sequentially. This property enables a much larger scale of experiments on common hardware. Moreover, to maximally utilize resources, actors can request fractional GPUs so that multiple actors can be co-located to the same GPU~\cite{lai2022fedscale}. Such a GPU-sharing technique accelerates FL optimization for lightweight models.

\myparatight{Scheduler and Synchronizer} They facilitate a lightweight and efficient mechanism for model parameter synchronization, worker group coordination, and task scheduling within a distributed training environment. They retrieve the training pipelines submitted by clients from a task queue, allocate them to suitable workers, and synchronize model parameters as required.

\myparatight{Object Store Memory} Following the distributed memory management of Ray, \ours allows storing and caching data objects during distributed computations using object store memory. Remote objects are cached in Ray's distributed shared-memory object store, and there is one object store per node in the cluster for easy access. 


By decoupling the execution layer from the application layer, clients remain unaware of the specific implementation details of the backend. As a result, users can concentrate solely on the application layer for implementing FL algorithms and submitting client training pipelines to the backend executor.

%% file: experiments.tex
\begin{figure*}[ht]
	\begin{subfigure}[b]{\linewidth}
		\centering
		\includegraphics[width=\linewidth]{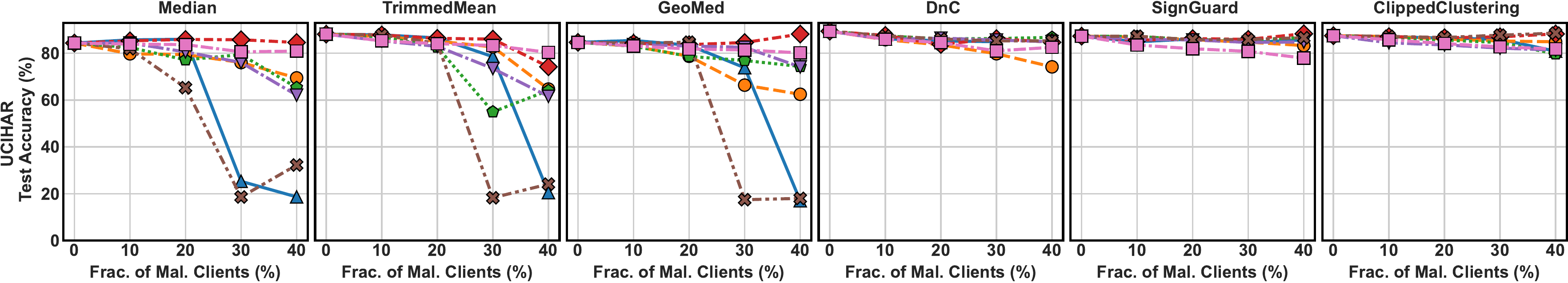}
	\end{subfigure}
    \begin{subfigure}[b]{\linewidth}
		\centering
		\includegraphics[width=\linewidth]{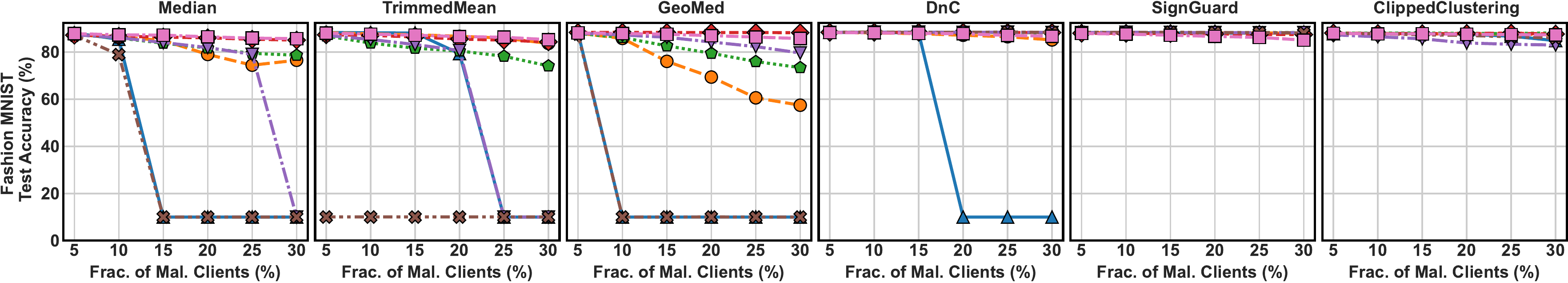}
	\end{subfigure}
        \vspace{0.2cm}
	\begin{subfigure}[b]{\linewidth}
		\centering
		\includegraphics[width=\linewidth]{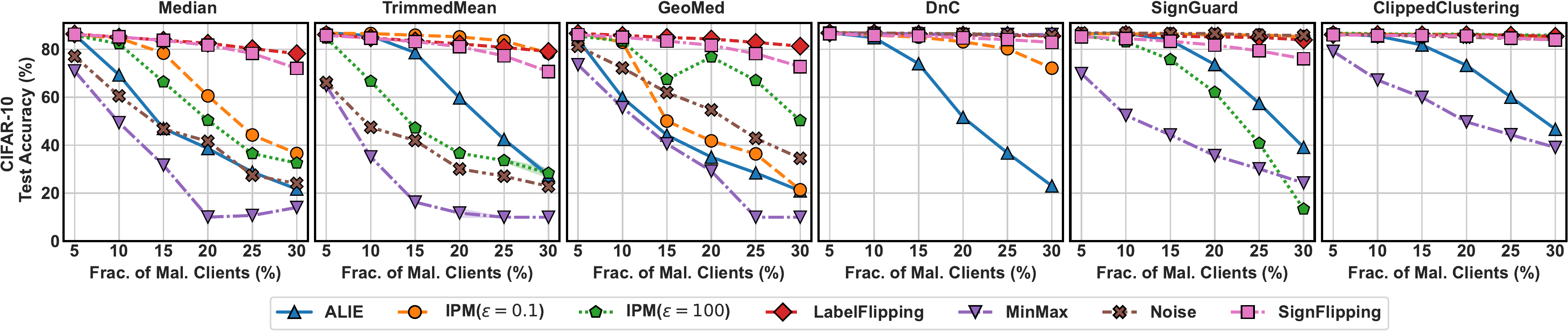}
	\end{subfigure}
	\vspace*{-0.7cm}
	\caption{\small Comparing AGRs under various attacks with IID partition. Simple attacks like LabelFlipping and SignFlipping are ineffective against most of the AGRs, while more advanced attacks such as ALIE and MinMax result in significant performance degradation across most settings, particularly as the fraction of malicious clients increases. Noticeably, traditional AGRs (i.e., \median~\cite{yin2018byzantine}, \tm~\cite{yin2018byzantine}, and \gm~\cite{pillutla2022robust}) exhibit vulnerabilities to several attacks, whereas advanced AGRs display greater resilience. }
	\label{atk_sim}
\end{figure*}

\begin{figure*}[ht]
	\begin{subfigure}[b]{\linewidth}
		\centering
		\includegraphics[width=\linewidth]{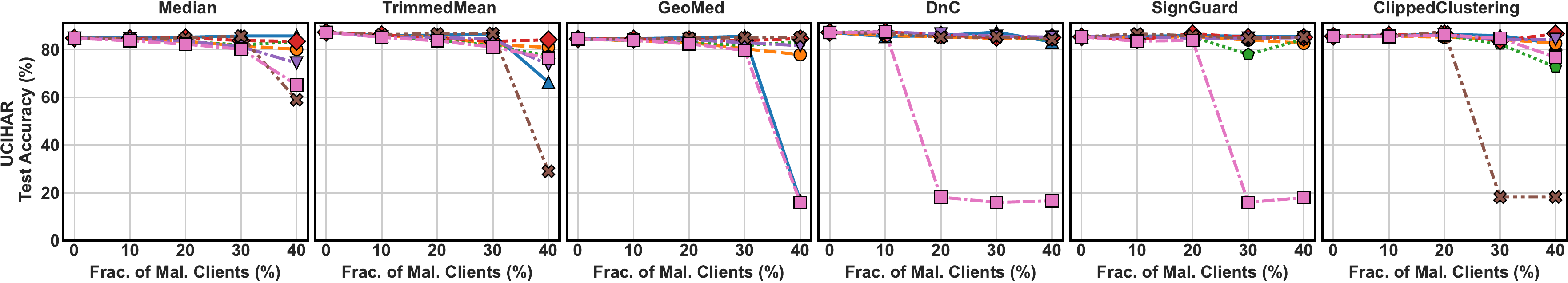}
	\end{subfigure}
  	\begin{subfigure}[b]{\linewidth}
		\centering
		\includegraphics[width=\linewidth]{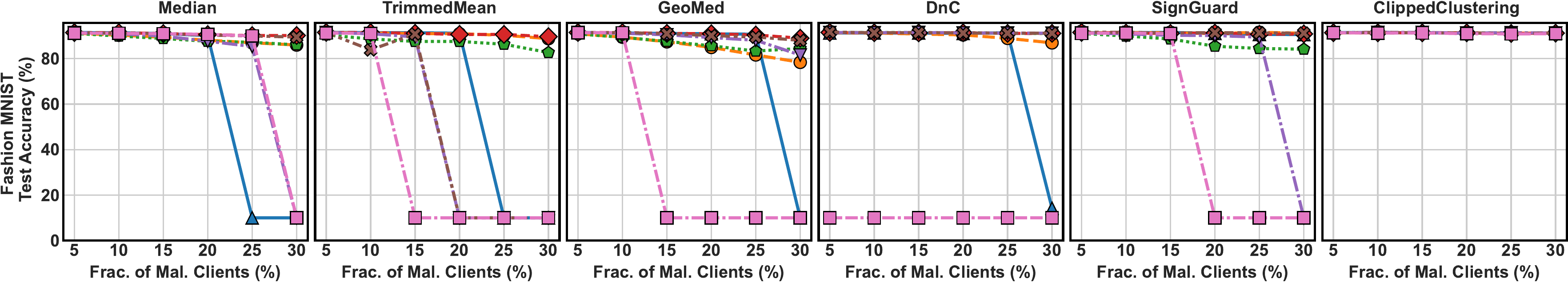}
	\end{subfigure}
        \vspace{0.2cm}
	\begin{subfigure}[b]{\linewidth}
		\centering
		\includegraphics[width=\linewidth]{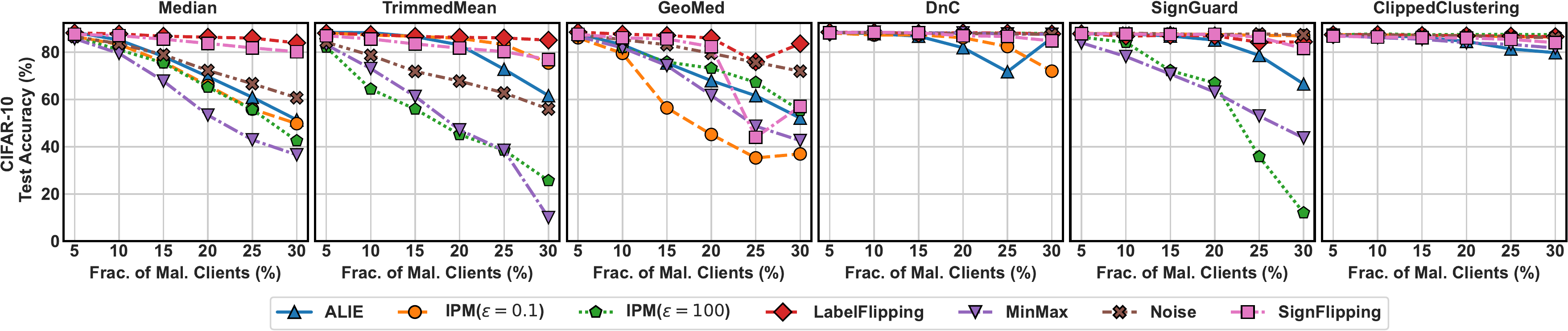}
	\end{subfigure}
	\vspace*{-.5cm}
	\caption{\small Evaluation of AGRs while utilizing 20 steps for local updates (i.e., $E_l=20$). \clippedclustering exhibits effective defense against most examined attacks with minimal performance loss, while others display distinct vulnerabilities depending on the attack type.}
	\label{atk_local_20}
\end{figure*}

\section{Experimental Evaluations}

In this section, we offer a comprehensive reassessment of six representative built-in AGRs, evaluating their performance against six attacks across three datasets with various settings and incorporating other relevant techniques. The results reveal novel insights in the area. Additionally, we test the scalability of \ours, with an emphasis on its adaptability to rising client numbers and computational demands.

\subsection{Datasets and Model Architectures}

\textbf{UCI-HAR}~\cite{uci_har} is an IoT dataset comprising observations from 30 volunteers carrying a waist-mounted smartphone and performing six distinct activities: walking, walking upstairs, walking downstairs, sitting, standing, and lying down. Each record is represented by 561 features, extracted from the time and frequency domain. With a total of 10,299 instances,  the dataset is naturally non-IID with data distributed across 30 clients. For this dataset, we utilize a CNN architecture, which includes two convolutional layers with 32 and 64 channels, respectively, both followed by ReLU activations and max-pooling operations. The network subsequently channels features through three fully connected layers with dimensions of 1024, 512, and the final output corresponding to the six classes.

\textbf{Fashion MNIST}~\cite{xiao2017fashion} consists of 50,000 gray-scale training samples and 10,000 test samples. It encompasses 10 different categories of clothing items, such as shoes, T-shirts, and dresses. The images in Fashion MNIST are of size $28 \times 28$. For this dataset, we employ a CNN with two convolutional layers. The first layer has 32 channels with a 3x3 kernel, and the second has 64 channels. Following the convolutional operations, the features are passed through three fully connected layers with 600, 120, and 10 neurons respectively. 

\textbf{CIFAR10}~\cite{krizhevsky2009learning} contains 50,000 color training samples and 10,000 test samples. It comprises color images of various objects classified into 10 categories, including airplanes and automobiles. The images in CIFAR10 have dimensions of $32 \times 32 \times 3$. For CIFAR10, we employ a modified ResNet architecture~\cite{He_2016_CVPR}, termed as ResNet10, which is shallower with only 10 layers. 
This design makes the model more suitable for resource-constrained IoT configurations while not significantly decreasing performance.

\subsection{Learning and Attack Settings}
We split Fashion MNIST and CIFAR10 into 60 distinct subsets and allocate them to 60 clients, utilizing both IID (independently and identically distributed) and non-IID strategies. For the IID approach, we assume homogeneity in data points, with each subset representing a random sampling of the entire dataset, ensuring statistical consistency. For the non-IID partition, we follow prior work~\cite{lin2020ensemble,li2023experimental} and model the non-IID data distributions with a Dirichlet distribution $\bm{p}_l \sim \textit{Dir}_K(\alpha)$, in which a smaller $\alpha$ indicates a stronger divergence from IID. Then we allocate a $\bm{p}_{l,k}$ proportion of the training samples of class $l$ to client $k$.

In our experiments, the models are trained for 2000 communication rounds using \fedsgd and 400 rounds for \fedavg. By default, the batch size is set to 64. For \fedsgd, we adopt the learning rates $\eta_l=1.0$ and $\eta_g=0.1$, with the latter undergoing a decay to 0.01 commencing from the 1501st round. Conversely, for \fedavg, we specify 20 local steps per round with parameters set as $\eta_l=0.1$, and $\eta_g=1.0$.

Additionally, in our assessment, we evaluate the six attacks detailed in Section~\ref{scope}. To rigorously stress-test the defense techniques, we introduce malicious clients, varying their proportion from 0\% to 40\% in the simulations.

\label{sec_exp}



\subsection{Comparison of AGRs under various settings}

\label{sec_agg}

For the sake of generality, we first examine selected AGRs with the standard \fedsgd and \fedavg. Fig.~\ref{atk_sim}  and Fig.~\ref{atk_local_20} depict the overall comparisons of different AGRs with respect to test accuracy under seven attack configurations. \textit{Overall, traditional and naive AGRs (i.e., \median~\cite{yin2018byzantine}, \tm~\cite{yin2018byzantine}, and \gm~\cite{pillutla2022robust}) exhibit significant vulnerabilities to several attacks, whereas hybrid strategies (i.e., \dnc~\cite{shejwalkar2021manipulating}, \clippedclustering~\cite{li2023experimental}, and \signguard~\cite{xu2022byzantine}) that integrate multiple techniques exhibit superior resilience against various attacks}. This aligns with previous studies~\cite{li2023experimental,shejwalkar2021manipulating}, which suggest that traditional AGRs, dependent on either dimensional-level filtering or optimization rooted in Euclidean distance, fall short in countering sophisticated attacks. However, our experimental evaluations have unveiled further insights that enrich our comprehension of Byzantine-resilient FL. In what follows, we elucidate these findings and present the crucial pieces of evidence supporting them.

\vspace*{-.5em}
\begin{mybox}
\paragraphb{(Finding 1)} 
\textit{The effectiveness of adversarial attacks is contingent upon a confluence of factors, including the choice of dataset, defensive countermeasure, and the specific FL algorithm employed.}
\end{mybox}


\begin{itemize}[leftmargin=0.4cm]

\item When examining the performance on the UCI-HAR and Fashion MNIST datasets, it becomes evident that the majority of AGRs retain a superior accuracy compared to their performance on CIFAR10. A plausible explanation for this observed distinction lies in the inherent characteristics of the datasets. Both UCI-HAR and Fashion MNIST, in comparison to CIFAR10, are considered to have relatively lower levels of complexity. As such, model updates trained on UCI-HAR and Fashion MNIST might face fewer challenges in diversity, leading to better performance under AGRs. 

\item The effectiveness of attacks varies significantly based on the AGR employed. As an example, for CIFAR10+\fedsgd, \median, \tm, \gm strategies show gradual declines in test accuracy as malicious clients increase, with MinMax and ALIE attacks being particularly effective. However, \dnc remains resilient against MinMax while still showing its vulnerability to ALIE. We note that both ALIE and MinMax leverage the variance of benign updates. However, MinMax typically amplifies the magnitudes to larger values, thereby presenting a mixed set of advantages and disadvantages. On the one hand, larger magnitudes have the potential to push the global model further away if they are not filtered out by the AGRs. On the other hand, these amplified magnitudes are more easily detectable by certain defense mechanisms, such as \dnc in our experiments. 


\item The SignFlipping attack, while exhibiting limited effectiveness under \fedsgd, leads to substantial performance degradation across various settings when integrated with \fedavg. Taking Fashion MNIST as an example, when combined with \fedsgd, SignFlipping makes little impact on all defenses, even with as many as 30\% malicious clients present. However, when associated with \fedavg, it disrupts four out of the six AGRs with just 15\% malicious clients in the mix. In contrast, both ALIE and MinMax show significant impacts with \fedsgd but underperform when integrated with \fedavg.


\end{itemize}

\vspace*{-.5em}
\begin{mybox}
\paragraphb{(Finding 2)} 
\textit{The robustness of defenses in existing studies may be overrated owing to their insufficiency in comprehensive evaluation under wide-ranging settings.}
\end{mybox}



\begin{itemize}[leftmargin=0.4cm]
    \item We notice that both \dnc and \signguard, which claimed robustness against several attacks with up to 20\% malicious clients for IID datasets, were evaluated with \fedsgd in their original works. However, the potential vulnerabilities were not detected when applied under the well-known algorithm \fedavg. Fig.~\ref{atk_local_20} show that they both become more vulnerable to SignFlipping while utilizing multiple (\eg $\eta_l=20$) steps for local updates. Surprisingly, \dnc cannot sustain as little as $5\%$ of malicious SignFlipping attack clients when on Fashion MNIST under \fedavg.
    \item In our experiments, \signguard effectively counters original ALIE and MinMax attacks by capitalizing on their distinct sign statistics. However, when \signguard is subjected to a more intensive examination in which we invert half of the signs—while maintaining the sign statistics (\ie the ratio of positive signs)—before computing the standard deviation for malicious updates, its robustness becomes compromised, making it susceptible to attacker bypass.
    \item We also evaluated two recent techniques, \cc~\cite{pmlrv139karimireddy21a} and \bucketing~\cite{karimireddy2022byzantinerobust} in our work. However, neither of these techniques could match the performance of other examined AGRs under similar experimental settings. We omitted the detailed results due to space limitations.
\end{itemize}

\subsection{Beyond AGRs: Additional Factors}

In addition to AGRs, we further examine a series of factors that may affect the robustness under adversarial attacks.

\vspace*{-.3em}
\begin{mybox}
\paragraphb{(Finding 3)} 
\textit{Additionally, various factors, including data heterogeneity, differential privacy (DP) noise, and momentum, exert considerable influence on the Byzantine resilience of defense strategies.}
\end{mybox}


\begin{figure}[t]
    \centering
    \includegraphics[width=0.96\linewidth]{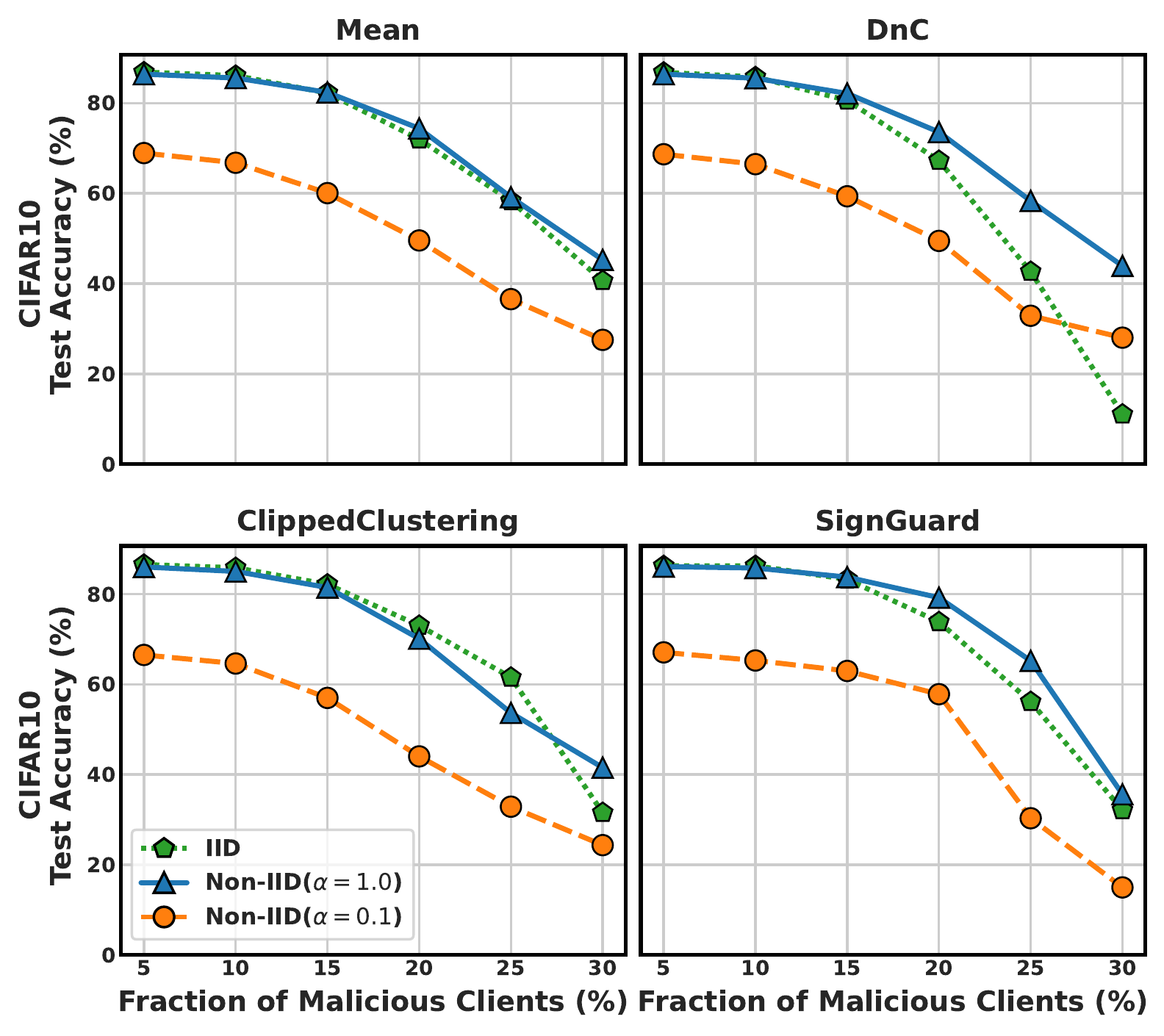}
    \caption{Impact of various degrees of non-IID on the robustness of AGRs against ALIE attack. A lower $\alpha$ value indicates a higher degree of non-IID.}
	\label{atk_noniid}
\end{figure}

\subsubsection{Impact of the degree of non-IID in datasets}
Fig.~\ref{atk_noniid} illustrates the test accuracy results of the ALIE attack on CIFAR10, varying the levels of non-IID partitioning. The figure shows that the effectiveness of the attack increases significantly when the dataset is highly non-IID  (e.g., $\alpha=0.1$). This is consistent with prior studies~\cite{fang2020local,shejwalkar2021manipulating,li2023experimental}. A commonly proposed explanation is that as the local data distributions become significantly different, the model updates diversify, thereby posing an additional challenge for AGRs to perform a proper aggregation.

\begin{figure}[t]
    \centering
    \includegraphics[width=0.96\linewidth]{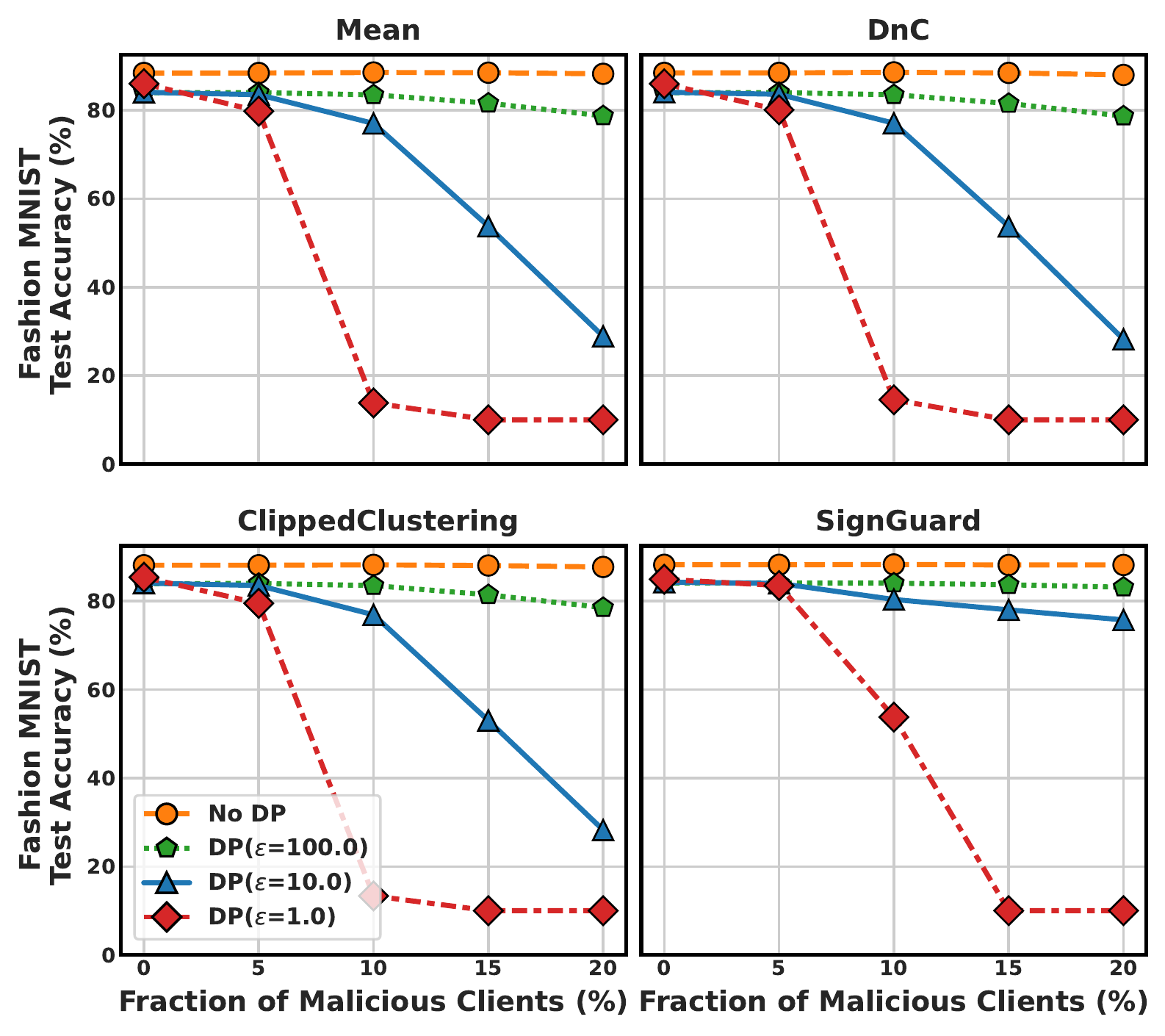}
    \caption{Impact of various DP levels on the robustness of AGRs against ALIE attack. A lower $\epsilon$ value indicates a lower budget for DP. All AGRs become more vulnerable to adversarial attacks with a lower privacy budget.}
	\label{atk_dp}
\end{figure}

\subsubsection{Impact of noise-adding for DP}
Next, we assess the influence of introducing Gaussian noise for DP on the resilience of defenses utilizing Fashion MNIST. Fig.~\ref{atk_dp} illustrates that as the privacy budget diminishes, the test accuracy across all AGRs declines more rapidly as the number of malicious clients increases. With a high budget (e.g., $\epsilon = 100.0$), all AGRs result in comparable accuracy levels to those without the incorporation of DP noise. In contrast, when the privacy parameter $\epsilon$ equals 1.0, the AGRs yield extremely low accuracy (nearly the same as random guessing) when $10\%$  of the clients are malicious.

\input{tables/momentum_table}
\subsubsection{Impact of Momentum} 

Momentum is considered a supplementary technique aiming at bolstering the robustness of Byzantine-resilient FL. To evaluate its effectiveness, we conduct experiments by training a ResNet10 on CIFAR10 under the ALIE attack. The obtained results are presented in Table~\ref{momentum}. Interestingly, the integration of server momentum demonstrates minimal enhancement in the robustness of the AGRs, particularly in situations involving a limited ratio (i.e., $10\%$) of malicious clients, while leading to lower accuracy in other situations. In contrast, the AGRs consistently exhibit significant improvements when client momentum is employed.


\subsubsection{The Risk of Gradient Explosion} 

\begin{figure}[t]
\centering
\includegraphics[width=\linewidth]{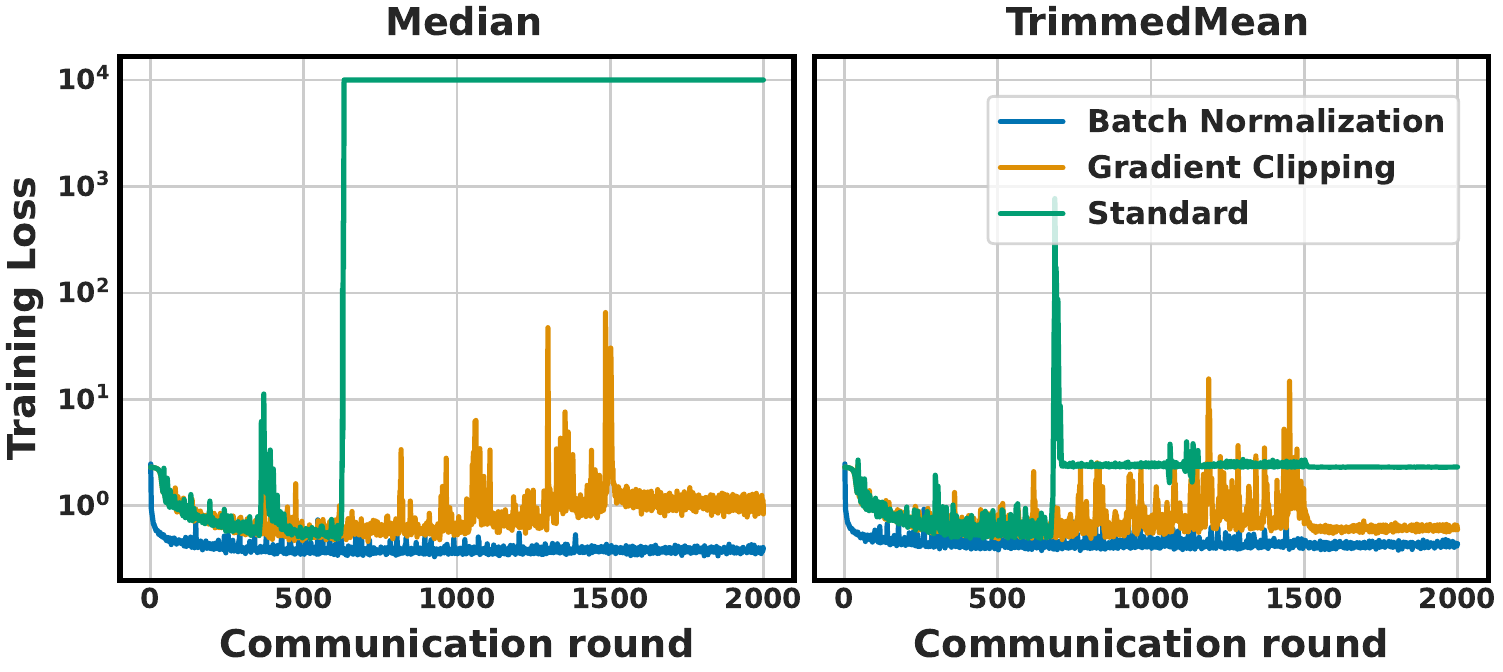}
\vspace{-.3cm}
\caption{Training loss on Fashion MNIST with noise attacks where $20\%$ of clients are malicious. The loss values are clamped to $[0, 10^4]$. The attacks lead to gradient explosions, which are mitigated by gradient clipping and batch normalization.}
\label{gradient_explosion}
\end{figure}

\label{sec_gradient_explosion}
Another interesting observation in Fig.~\ref{atk_sim} is that four AGRs fail to handle Noise attacks on Fashion MNIST even when only 10\% of the clients are malicious. Upon closer examination of the gradients and loss values, we observe significant gradient explosions among the benign clients, which prevent the model from converging to the optimal solution. We believe that the decline in performance is attributed not only to substantial deviations but also to the potential detrimental impact of gradient explosions caused by malicious attacks.

To validate our findings, we employ two measures to mitigate gradient explosions, i.e., gradient clipping and batch normalization, and show the comparison in Fig.~\ref{gradient_explosion}. When no measures are taken, the training loss shows sudden spikes and ends up with a large value, indicating that the model cannot converge to an appropriate solution. With gradient clipping and batch normalization, the training process becomes more stable. Batch normalization not only prevents gradient explosions but also accelerates the training process, facilitating faster convergence.

%% file: tables/momentum_table.tex
\begin{table*}[!t]
\caption{\small Test accuracy of CIFAR10 on different momentum settings. The numbers with the highest accuracy are in bold. Client momentum significantly improves the robustness.}
	\vspace*{-.4cm}
	\begin{center}
	\setlength\extrarowheight{2pt}
	\begin{adjustbox}{width=.95\linewidth,center}
        \begin{tabular}{c|ccc|ccc|ccc}
		    \hline
			  \multirow{3}{*}{\textbf{AGR}}  & \multicolumn{3}{c|}{10\% malicious clients} & \multicolumn{3}{c|}{15\% malicious clients} & \multicolumn{3}{c}{20\% malicious clients}\\
			 \cline{2-10}
			  & \makecell{No \\ Momentum} & \makecell{Server \\ Momentum} & \makecell{Client \\ Momentum} & \makecell{No \\ Momentum} & \makecell{Server \\ Momentum} & \makecell{Client \\ Momentum} & \makecell{No \\ Momentum} & \makecell{Server \\ Momentum} & \makecell{Client \\ Momentum}\\	
   \hline
   \hline
\mean~\cite{mcmahan2017communication} & 86.05 (0.20) & 88.46 (0.24) & \textbf{89.51 (0.22)} & 82.89 (0.46) & 80.72 (0.53) & \textbf{88.18 (0.34)} & 73.50 (0.91) & 59.92 (1.36) & \textbf{85.88 (0.22)}\\
\dnc~\cite{shejwalkar2021manipulating} & 85.84 (0.19) & 87.92 (0.45) & \textbf{89.40 (0.15)} & 81.45 (0.29) & 76.58 (0.08) & \textbf{87.89 (0.45)} & 68.63 (0.40) & 50.04 (3.88) & \textbf{83.60 (0.40)}\\
\median~\cite{yin2018byzantine} & 69.38 (0.58) & 48.39 (2.41) & \textbf{83.95 (0.31)} & 45.95 (3.05) & 27.52 (3.97) & \textbf{68.20 (0.63)} & 27.85 (10.7) & 14.29 (6.23) & \textbf{52.22 (3.73)}\\
\tm~\cite{yin2018byzantine} & 85.98 (0.29) & 87.81 (0.14) & \textbf{89.57 (0.18)} & 79.98 (0.47) & 72.40 (0.49) & \textbf{87.47 (0.12)} & 66.69 (0.47) & 41.32 (4.75) & \textbf{81.72 (0.44)}\\
\clippedclustering~\cite{li2023experimental} & 85.76 (0.16) & 87.86 (0.19) & \textbf{89.47 (0.20)} & 82.25 (0.25) & 73.23 (1.23) & \textbf{87.94 (0.36)} & 73.19 (0.98) & 41.52 (5.63) & \textbf{83.06 (0.62)}\\
			\hline
		\end{tabular}
	\end{adjustbox}
	\end{center}
	\label{momentum}
\end{table*}

%% file: related_work.tex

%% file: main.bbl
\begin{thebibliography}{10}
\providecommand{\url}[1]{#1}
\csname url@samestyle\endcsname
\providecommand{\newblock}{\relax}
\providecommand{\bibinfo}[2]{#2}
\providecommand{\BIBentrySTDinterwordspacing}{\spaceskip=0pt\relax}
\providecommand{\BIBentryALTinterwordstretchfactor}{4}
\providecommand{\BIBentryALTinterwordspacing}{\spaceskip=\fontdimen2\font plus
\BIBentryALTinterwordstretchfactor\fontdimen3\font minus
  \fontdimen4\font\relax}
\providecommand{\BIBforeignlanguage}[2]{{%
\expandafter\ifx\csname l@#1\endcsname\relax
\typeout{** WARNING: IEEEtran.bst: No hyphenation pattern has been}%
\typeout{** loaded for the language `#1'. Using the pattern for}%
\typeout{** the default language instead.}%
\else
\language=\csname l@#1\endcsname
\fi
#2}}
\providecommand{\BIBdecl}{\relax}
\BIBdecl

\bibitem{mcmahan2017communication}
B.~McMahan, E.~Moore, D.~Ramage, S.~Hampson, and B.~A. y~Arcas,
  ``Communication-efficient learning of deep networks from decentralized
  data,'' in \emph{Artificial intelligence and statistics}.\hskip 1em plus
  0.5em minus 0.4em\relax PMLR, 2017.

\bibitem{konevcny2015federated}
J.~Kone{\v{c}}n{\`y}, B.~McMahan, and D.~Ramage, ``Federated optimization:
  Distributed optimization beyond the datacenter,'' \emph{arXiv preprint
  arXiv:1511.03575}, 2015.

\bibitem{khan2021federated}
L.~U. Khan, W.~Saad, Z.~Han, E.~Hossain, and C.~S. Hong, ``Federated learning
  for internet of things: Recent advances, taxonomy, and open challenges,''
  \emph{IEEE Communications Surveys \& Tutorials}, vol.~23, no.~3, pp.
  1759--1799, 2021.

\bibitem{li2021byzantine}
S.~Li, E.~Ngai, and T.~Voigt, ``Byzantine-robust aggregation in federated
  learning empowered industrial iot,'' \emph{IEEE Transactions on Industrial
  Informatics}, vol.~19, no.~2, pp. 1165--1175, 2023.

\bibitem{arjevani2015communication}
Y.~Arjevani and O.~Shamir, ``Communication complexity of distributed convex
  learning and optimization,'' \emph{Advances in neural information processing
  systems}, vol.~28, 2015.

\bibitem{konevcny2016federated}
J.~Kone{\v{c}}n{\`y}, H.~B. McMahan, F.~X. Yu, P.~Richt{\'a}rik, A.~T. Suresh,
  and D.~Bacon, ``Federated learning: Strategies for improving communication
  efficiency,'' \emph{arXiv preprint arXiv:1610.05492}, 2016.

\bibitem{lyu2020threats}
L.~Lyu, H.~Yu, and Q.~Yang, ``Threats to federated learning: A survey,''
  \emph{arXiv preprint arXiv:2003.02133}, 2020.

\bibitem{rodriguez2023survey}
N.~Rodr{\'\i}guez-Barroso, D.~Jim{\'e}nez-L{\'o}pez, M.~V. Luz{\'o}n,
  F.~Herrera, and E.~Mart{\'\i}nez-C{\'a}mara, ``Survey on federated learning
  threats: Concepts, taxonomy on attacks and defences, experimental study and
  challenges,'' \emph{Information Fusion}, vol.~90, pp. 148--173, 2023.

\bibitem{li2023experimental}
S.~Li, E.~C.-H. Ngai, and T.~Voigt, ``An experimental study of byzantine-robust
  aggregation schemes in federated learning,'' \emph{IEEE Transactions on Big
  Data}, 2023.

\bibitem{jere2020taxonomy}
M.~S. Jere, T.~Farnan, and F.~Koushanfar, ``A taxonomy of attacks on federated
  learning,'' \emph{IEEE Security \& Privacy}, vol.~19, no.~2, 2020.

\bibitem{xie2019dba}
C.~Xie, K.~Huang, P.-Y. Chen, and B.~Li, ``Dba: Distributed backdoor attacks
  against federated learning,'' in \emph{International Conference on Learning
  Representations}, 2020.

\bibitem{bagdasaryan2020backdoor}
E.~Bagdasaryan, A.~Veit, Y.~Hua, D.~Estrin, and V.~Shmatikov, ``How to backdoor
  federated learning,'' in \emph{International Conference on Artificial
  Intelligence and Statistics}.\hskip 1em plus 0.5em minus 0.4em\relax PMLR,
  2020, pp. 2938--2948.

\bibitem{andreina2021baffle}
S.~Andreina, G.~A. Marson, H.~M{\"o}llering, and G.~Karame, ``Baffle: Backdoor
  detection via feedback-based federated learning,'' in \emph{2021 IEEE 41st
  International Conference on Distributed Computing Systems (ICDCS)}.\hskip 1em
  plus 0.5em minus 0.4em\relax IEEE, 2021, pp. 852--863.

\bibitem{fang2020local}
M.~Fang, X.~Cao, J.~Jia, and N.~Gong, ``Local model poisoning attacks to
  byzantine-robust federated learning,'' in \emph{29th USENIX Security
  Symposium (USENIX Security 20)}, 2020, pp. 1605--1622.

\bibitem{blanchard2017machine}
P.~Blanchard, E.~M. El~Mhamdi, R.~Guerraoui, and J.~Stainer, ``Machine learning
  with adversaries: Byzantine tolerant gradient descent,'' in \emph{Proceedings
  of the 31st International Conference on Neural Information Processing
  Systems}, 2017.

\bibitem{yin2018byzantine}
D.~Yin, Y.~Chen, R.~Kannan, and P.~Bartlett, ``Byzantine-robust distributed
  learning: Towards optimal statistical rates,'' in \emph{International
  Conference on Machine Learning}.\hskip 1em plus 0.5em minus 0.4em\relax PMLR,
  2018.

\bibitem{chen2017distributed}
Y.~Chen, L.~Su, and J.~Xu, ``Distributed statistical machine learning in
  adversarial settings: Byzantine gradient descent,'' \emph{Proceedings of the
  ACM on Measurement and Analysis of Computing Systems}, 2017.

\bibitem{pmlrv139karimireddy21a}
S.~P. Karimireddy, L.~He, and M.~Jaggi, ``Learning from history for byzantine
  robust optimization,'' in \emph{Proceedings of the 38th International
  Conference on Machine Learning}, ser. Proceedings of Machine Learning
  Research, M.~Meila and T.~Zhang, Eds., vol. 139.\hskip 1em plus 0.5em minus
  0.4em\relax PMLR, 2021.

\bibitem{sattler2020byzantine}
F.~Sattler, K.-R. M{\"u}ller, T.~Wiegand, and W.~Samek, ``On the byzantine
  robustness of clustered federated learning,'' in \emph{ICASSP 2020-2020 IEEE
  International Conference on Acoustics, Speech and Signal Processing
  (ICASSP)}.\hskip 1em plus 0.5em minus 0.4em\relax IEEE, 2020, pp. 8861--8865.

\bibitem{9833647}
V.~Shejwalkar, A.~Houmansadr, P.~Kairouz, and D.~Ramage, ``Back to the drawing
  board: A critical evaluation of poisoning attacks on production federated
  learning,'' in \emph{2022 IEEE Symposium on Security and Privacy (SP)}, 2022,
  pp. 1354--1371.

\bibitem{hu2021challenges}
S.~Hu, J.~Lu, W.~Wan, and L.~Y. Zhang, ``Challenges and approaches for
  mitigating byzantine attacks in federated learning,'' \emph{arXiv preprint
  arXiv:2112.14468}, 2021.

\bibitem{baruch2019little}
G.~Baruch, M.~Baruch, and Y.~Goldberg, ``A little is enough: Circumventing
  defenses for distributed learning,'' \emph{Advances in Neural Information
  Processing Systems}, vol.~32, 2019.

\bibitem{xie2020fall}
C.~Xie, O.~Koyejo, and I.~Gupta, ``Fall of empires: Breaking byzantine-tolerant
  sgd by inner product manipulation,'' in \emph{Uncertainty in Artificial
  Intelligence}.\hskip 1em plus 0.5em minus 0.4em\relax PMLR, 2020, pp.
  261--270.

\bibitem{kairouz2021advances}
P.~Kairouz, H.~B. McMahan, B.~Avent, A.~Bellet, M.~Bennis, A.~N. Bhagoji,
  K.~Bonawitz, Z.~Charles, G.~Cormode, R.~Cummings \emph{et~al.}, ``Advances
  and open problems in federated learning,'' \emph{Foundations and
  Trends{\textregistered} in Machine Learning}, vol.~14, no. 1--2, pp. 1--210,
  2021.

\bibitem{khan2023pitfalls}
M.~A. Khan, V.~Shejwalkar, A.~Houmansadr, and F.~M. Anwar, ``On the pitfalls of
  security evaluation of robust federated learning,'' in \emph{2023 IEEE
  Security and Privacy Workshops (SPW)}.\hskip 1em plus 0.5em minus 0.4em\relax
  IEEE, 2023, pp. 57--68.

\bibitem{reddi2020adaptive}
S.~Reddi, Z.~Charles, M.~Zaheer, Z.~Garrett, K.~Rush, J.~Kone{\v{c}}n{\`y},
  S.~Kumar, and H.~B. McMahan, ``Adaptive federated optimization,'' \emph{arXiv
  preprint arXiv:2003.00295}, 2020.

\bibitem{ju2023accelerating}
L.~Ju, T.~Zhang, S.~Toor, and A.~Hellander, ``Accelerating fair federated
  learning: Adaptive federated adam,'' \emph{arXiv preprint arXiv:2301.09357},
  2023.

\bibitem{liconvergence}
X.~Li, K.~Huang, W.~Yang, S.~Wang, and Z.~Zhang, ``On the convergence of fedavg
  on non-iid data,'' in \emph{International Conference on Learning
  Representations}, 2020.

\bibitem{shejwalkar2021manipulating}
V.~Shejwalkar and A.~Houmansadr, ``Manipulating the byzantine: Optimizing model
  poisoning attacks and defenses for federated learning,'' in \emph{NDSS},
  2021.

\bibitem{li2019rsa}
L.~Li, W.~Xu, T.~Chen, G.~B. Giannakis, and Q.~Ling, ``Rsa: Byzantine-robust
  stochastic aggregation methods for distributed learning from heterogeneous
  datasets,'' in \emph{Proceedings of the AAAI Conference on Artificial
  Intelligence}, vol.~33, no.~01, 2019, pp. 1544--1551.

\bibitem{cao2021fltrust}
X.~Cao, M.~Fang, J.~Liu, and N.~Z. Gong, ``Fltrust: Byzantine-robust federated
  learning via trust bootstrapping,'' in \emph{ISOC Network and Distributed
  System Security Symposium (NDSS)}, 2021.

\bibitem{9887909}
C.~Xu, Y.~Jia, L.~Zhu, C.~Zhang, G.~Jin, and K.~Sharif, ``Tdfl: Truth discovery
  based byzantine robust federated learning,'' \emph{IEEE Transactions on
  Parallel and Distributed Systems}, vol.~33, no.~12, 2022.

\bibitem{sageflow}
J.~Park, D.-J. Han, M.~Choi, and J.~Moon, ``Sageflow: Robust federated learning
  against both stragglers and adversaries,'' in \emph{Advances in Neural
  Information Processing Systems}, M.~Ranzato, A.~Beygelzimer, Y.~Dauphin,
  P.~Liang, and J.~W. Vaughan, Eds., vol.~34.\hskip 1em plus 0.5em minus
  0.4em\relax Curran Associates, Inc., 2021, pp. 840--851.

\bibitem{gorbunov2023variance}
E.~Gorbunov, S.~Horv{\'a}th, P.~Richt{\'a}rik, and G.~Gidel, ``Variance
  reduction is an antidote to byzantines: Better rates, weaker assumptions and
  communication compression as a cherry on the top,'' in \emph{International
  Conference on Learning Representations}, 2023.

\bibitem{9153949}
Z.~Wu, Q.~Ling, T.~Chen, and G.~B. Giannakis, ``Federated variance-reduced
  stochastic gradient descent with robustness to byzantine attacks,''
  \emph{IEEE Transactions on Signal Processing}, vol.~68, pp. 4583--4596, 2020.

\bibitem{pillutla2022robust}
K.~Pillutla, S.~M. Kakade, and Z.~Harchaoui, ``Robust aggregation for federated
  learning,'' \emph{IEEE Transactions on Signal Processing}, vol.~70, pp.
  1142--1154, 2022.

\bibitem{xu2022byzantine}
J.~Xu, S.-L. Huang, L.~Song, and T.~Lan, ``Byzantine-robust federated learning
  through collaborative malicious gradient filtering,'' in \emph{2022 IEEE 42nd
  International Conference on Distributed Computing Systems (ICDCS)}.\hskip 1em
  plus 0.5em minus 0.4em\relax IEEE, 2022, pp. 1223--1235.

\bibitem{alistarh2018byzantine}
D.~Alistarh, Z.~Allen-Zhu, and J.~Li, ``Byzantine stochastic gradient
  descent,'' \emph{Advances in Neural Information Processing Systems}, 2018.

\bibitem{yin2021comprehensive}
X.~Yin, Y.~Zhu, and J.~Hu, ``A comprehensive survey of privacy-preserving
  federated learning: A taxonomy, review, and future directions,'' \emph{ACM
  Computing Surveys (CSUR)}, vol.~54, no.~6, pp. 1--36, 2021.

\bibitem{li2020review}
L.~Li, Y.~Fan, M.~Tse, and K.-Y. Lin, ``A review of applications in federated
  learning,'' \emph{Computers \& Industrial Engineering}, vol. 149, p. 106854,
  2020.

\bibitem{bernstein2018signsgd}
J.~Bernstein, J.~Zhao, K.~Azizzadenesheli, and A.~Anandkumar, ``signsgd with
  majority vote is communication efficient and fault tolerant,'' \emph{arXiv
  preprint arXiv:1810.05291}, 2018.

\bibitem{damaskinos2019aggregathor}
G.~Damaskinos, E.-M. El-Mhamdi, R.~Guerraoui, A.~Guirguis, and S.~Rouault,
  ``Aggregathor: Byzantine machine learning via robust gradient aggregation,''
  \emph{Proceedings of Machine Learning and Systems}, vol.~1, pp. 81--106,
  2019.

\bibitem{han2023fedmlsecurity}
S.~Han, B.~Buyukates, Z.~Hu, H.~Jin, W.~Jin, L.~Sun, X.~Wang, C.~Xie, K.~Zhang
  \emph{et~al.}, ``Fedmlsecurity: A benchmark for attacks and defenses in
  federated learning and llms,'' \emph{arXiv preprint arXiv:2306.04959}, 2023.

\bibitem{lai2022fedscale}
F.~Lai, Y.~Dai, S.~Singapuram, J.~Liu, X.~Zhu, H.~Madhyastha, and M.~Chowdhury,
  ``Fedscale: Benchmarking model and system performance of federated learning
  at scale,'' in \emph{International Conference on Machine Learning}.\hskip 1em
  plus 0.5em minus 0.4em\relax PMLR, 2022, pp. 11\,814--11\,827.

\bibitem{yao2022benchmark}
L.~Yao, D.~Gao, Z.~Wang, Y.~Xie, W.~Kuang, D.~Chen, H.~Wang, C.~Dong, B.~Ding,
  and Y.~Li, ``A benchmark for federated hetero-task learning,'' \emph{arXiv
  preprint arXiv}, vol. 2206, 2022.

\bibitem{chen2022pfl}
D.~Chen, D.~Gao, W.~Kuang, Y.~Li, and B.~Ding, ``pfl-bench: A comprehensive
  benchmark for personalized federated learning,'' \emph{Advances in Neural
  Information Processing Systems}, vol.~35, pp. 9344--9360, 2022.

\bibitem{mugunthan2020privacyfl}
V.~Mugunthan, A.~Peraire-Bueno, and L.~Kagal, ``Privacyfl: A simulator for
  privacy-preserving and secure federated learning,'' in \emph{Proceedings of
  the 29th ACM International Conference on Information \& Knowledge
  Management}, 2020, pp. 3085--3092.

\bibitem{ziller2021pysyft}
A.~Ziller, A.~Trask, A.~Lopardo, B.~Szymkow, B.~Wagner, E.~Bluemke, J.-M.
  Nounahon, J.~Passerat-Palmbach, K.~Prakash, N.~Rose \emph{et~al.}, ``Pysyft:
  A library for easy federated learning,'' in \emph{Federated Learning
  Systems}.\hskip 1em plus 0.5em minus 0.4em\relax Springer, 2021, pp.
  111--139.

\bibitem{qin2023revisiting}
Z.~Qin, L.~Yao, D.~Chen, Y.~Li, B.~Ding, and M.~Cheng, ``Revisiting
  personalized federated learning: Robustness against backdoor attacks,''
  \emph{arXiv preprint arXiv:2302.01677}, 2023.

\bibitem{li2020federated}
T.~Li, A.~K. Sahu, M.~Zaheer, M.~Sanjabi, A.~Talwalkar, and V.~Smith,
  ``Federated optimization in heterogeneous networks,'' \emph{Proceedings of
  Machine learning and systems}, vol.~2, pp. 429--450, 2020.

\bibitem{wang2020tackling}
J.~Wang, Q.~Liu, H.~Liang, G.~Joshi, and H.~V. Poor, ``Tackling the objective
  inconsistency problem in heterogeneous federated optimization,''
  \emph{Advances in neural information processing systems}, vol.~33, 2020.

\bibitem{he2020fedml}
C.~He, S.~Li, J.~So, X.~Zeng, M.~Zhang, H.~Wang, X.~Wang, P.~Vepakomma,
  A.~Singh, H.~Qiu \emph{et~al.}, ``Fedml: A research library and benchmark for
  federated machine learning,'' \emph{arXiv preprint arXiv:2007.13518}, 2020.

\bibitem{liaw2018tune}
R.~Liaw, E.~Liang, R.~Nishihara, P.~Moritz, J.~E. Gonzalez, and I.~Stoica,
  ``Tune: A research platform for distributed model selection and training,''
  \emph{arXiv preprint arXiv:1807.05118}, 2018.

\bibitem{moritz2018ray}
P.~Moritz, R.~Nishihara, S.~Wang, A.~Tumanov, R.~Liaw, E.~Liang, M.~Elibol,
  Z.~Yang, W.~Paul, M.~I. Jordan \emph{et~al.}, ``Ray: A distributed framework
  for emerging $\{$AI$\}$ applications,'' in \emph{13th USENIX Symposium on
  Operating Systems Design and Implementation (OSDI 18)}, 2018.

\bibitem{caldas2018leaf}
S.~Caldas, S.~M.~K. Duddu, P.~Wu, T.~Li, J.~Kone{\v{c}}n{\`y}, H.~B. McMahan,
  V.~Smith, and A.~Talwalkar, ``Leaf: A benchmark for federated settings,''
  \emph{arXiv preprint arXiv:1812.01097}, 2018.

\bibitem{JMLR:v24:22-0440}
D.~Zeng, S.~Liang, X.~Hu, H.~Wang, and Z.~Xu, ``Fedlab: A flexible federated
  learning framework,'' \emph{Journal of Machine Learning Research}, vol.~24,
  no. 100, pp. 1--7, 2023.

\bibitem{holt1972some}
R.~C. Holt, ``Some deadlock properties of computer systems,'' \emph{ACM
  Computing Surveys (CSUR)}, vol.~4, no.~3, pp. 179--196, 1972.

\bibitem{uci_har}
D.~Anguita, A.~Ghio, L.~Oneto, X.~Parra, J.~Reyes-Ortiz \emph{et~al.}, ``A
  public domain dataset for human activity recognition using smartphones,'' in
  \emph{21th European Symposium on Artificial Neural Networks, Computational
  Intelligence and Machine Learning (ESANN)}.\hskip 1em plus 0.5em minus
  0.4em\relax CIACO, 2013.

\bibitem{xiao2017fashion}
H.~Xiao, K.~Rasul, and R.~Vollgraf, ``Fashion-mnist: a novel image dataset for
  benchmarking machine learning algorithms,'' \emph{arXiv preprint
  arXiv:1708.07747}, 2017.

\bibitem{krizhevsky2009learning}
A.~Krizhevsky, G.~Hinton \emph{et~al.}, ``Learning multiple layers of features
  from tiny images,'' 2009.

\bibitem{He_2016_CVPR}
K.~He, X.~Zhang, S.~Ren, and J.~Sun, ``Deep residual learning for image
  recognition,'' in \emph{Proceedings of the IEEE Conference on Computer Vision
  and Pattern Recognition (CVPR)}, June 2016.

\bibitem{lin2020ensemble}
T.~Lin, L.~Kong, S.~U. Stich, and M.~Jaggi, ``Ensemble distillation for robust
  model fusion in federated learning,'' \emph{Advances in Neural Information
  Processing Systems}, vol.~33, pp. 2351--2363, 2020.

\bibitem{karimireddy2022byzantinerobust}
S.~P. Karimireddy, L.~He, and M.~Jaggi, ``Byzantine-robust learning on
  heterogeneous datasets via bucketing,'' in \emph{International Conference on
  Learning Representations}, 2022.

\end{thebibliography}
